\documentstyle[aps,epsfig,prb]{revtex}

\newcommand{\be}{\begin{eqnarray}}
\newcommand{\en}{\end{eqnarray}}

\newcommand{\up}{\uparrow}
\newcommand{\down}{\downarrow}

\begin{document}
\draft

\title{New  Collective Modes in the Superconducting Ground States in the
Gauge Theory Description of the Cuprates}

\author{Patrick A. Lee}

\address{Department of Physics,
Massachusetts Institute of Technology,
Cambridge, Massachusetts 02139}

\author{Naoto Nagaosa}

\address{Department of Applied Physics,
University of Tokyo,
7--3--1 Hongo, Bunkyo-ku, Tokyo 113, Japan }

\date{\today}
\maketitle

\begin{abstract}
In the slave boson mean field treatment of the $t$-$J$ model, the ground
state for small doping is a $d$-wave superconductor.  A conventional
superconductor has collective modes associated with the amplitude and
phase of the pairing order parameter.  Here, the hopping matrix element
$\chi_{ij}$ is also a mean field order parameter.  We therefore expect
that new collective modes will be introduced with these additional complex
degrees of freedom.  We compute the new collective modes and their
spectral functions by numerically diagonalizing the matrix which
describes the fluctuations about the mean field solution.  We also show
that the SU(2) gauge theory formulation allows us to predict and classify
these collective modes.  Indeed, the SU(2) formulation is essential in
order to avoid spurious collective modes as the doping $x$ goes to zero.
The most important new collective modes are the $\theta$ mode and the
longitudinal and transverse $\phi$ gauge modes.  The $\theta$ mode
corresponds to fluctuations of the staggered flux phase and creates
orbital current fluctuations.  The $\phi$ gauge modes correspond to
out-of-phase fluctuations of the {\em amplitudes} of the pairing and the
hopping matrix elements.  We compute the neutron scattering cross-section
which couples to the $\theta$ mode and inelastic X-ray scattering
cross-section which couples to the fluctuation in  the
real part of $\chi_{ij}$.
  In addition, we show that the latter
fluctuation at $(\pi,\pi)$ may be coupled to the buckling
phonon mode in the LTO phase of LSCO and may be detected optically.
Experimental searches of these collective modes will
serve as important tests of this line of attack on the high T$_c$ problem.
\end{abstract}
\vskip 0.5 in

\baselineskip15pt

\section{Introduction}
It is now a widely accepted view that the problem of high T$_c$
superconductivity is that of doping into a Mott
insulator.  The simplest model which captures the physics of the
strong correlation inherent in this problem is the $t$-$J$
model. The no-double-occupant constraint leads naturally to a gauge
theory. The mean field decoupling of this theory is the
formal language which describes Anderson's physical idea of a
resonating valence bond (RVB).\cite{1}  The mean field theory indeed
enjoys a number of successes.  Notably it predicted the appearance of
$d$-wave superconductivity and the existence of a spin
gap phase in the underdoped region.\cite{2,3,4}  The properties of
this spin gap phase are in remarkable agreement with those  of the
pseudogap phenomenology.  Most striking among these is that the
quasiparticle spectral function becomes a coherent peak with
small spectral weight.  This
property emerges naturally out of the mean field picture
of the condensation of bosons.

Despite these successes, the theory has not enjoyed wide following
because the gauge fluctuations are strong and the theory
does not have a well controlled small expansion parameter, except
formal ones such as the large $N$ expansion.  For example,
Nayak\cite{5} has raised a number of objections on general grounds.  However,
we believe these general arguments have been adequately
answered in the ensuing comments and discussions.\cite{6,7,8,sud}  
Here we briefly summarize our point of view.

Initially, the gauge field has infinite coupling in order to enforce
the constraint on a lattice scale.  Upon integrating out
high energy fermionic degrees of freedom, the coupling constant
becomes of order unity.  Then there is a chance that the mean
field theory and fluctuations about it may be qualitatively correct
at intermediate temperature scales.  At low energy scales,
nonperturbative effects related to compact gauge theories may come
in, giving rise to the phenomenon of confinement.  This
phenomenon is well known at half-filling.   The mean field solution
gives a $\pi$-flux state with Dirac spectrum centered about
$\left( \pm {\pi \over 2},\pm {\pi \over 2 } \right)$.  Coupling with
compact U(1) gauge field leads to confinement and chiral
symmetry breaking in particle physics language, which is equivalent
to N\'{e}el ordering.\cite{9}  The idea is that with doping, the
appearance of dissipation suppresses confinement,\cite{10} and the N\'{e}el
state is rapidly destroyed.  Then the mean field
description may be qualitatively correct beyond a small, but finite,
doping concentration.  At low temperatures characterized
by the boson condensation scale $xt$, superconductivity emerges.  In
this theory the superconducting state is described as the
Higgs phase associated with bose condensation.  (We prefer to refer
to this as the Higgs phase rather than bose condensation
because the bose field is not gauge invariant.)  However, here lies
one of the significant failures of the weak coupling gauge
theory description.  As long as the gauge fluctuation is treated as
Gaussian, the Ioffe-Larkin law holds and one predicts that
the superfluid density $\rho_s(T)$ behaves as $\rho_s(T) \approx ax -
bx^2T$.  The $ax$ term agrees with experiment while the
$-bx^2T$ term does not.\cite{11,12}  This failure is traced to the fact that in
the Gaussian approximation, the current carried by the
quasiparticles in the superconducting state is proportional to
$xv_F$.  We believe this failure is a sign that nonperturbative
effects again become important and confinement takes place, so that
the low energy quasiparticles near the nodes behave
like BCS quasiparticles which carry the full current $v_F$.  For the
pure gauge theory, the confinement always occurs in (2+1)D
however small the coupling constant of the gauge field is.
When the bose field is coupled to the gauge field, i.e., Higgs model,
it is known that the confinement phase are smoothly connected to each
other.\cite{13}
Recently, we showed that in (2+1)D Higgs model the phase should be
considered confined everywhere.\cite{14}  Thus the appearance of
confinement at low energy scale is not surprising, but the dichotomy
between the success of the bose condensation
phenomenology on the one hand (in explaining the quasiparticle
spectral weight mentioned earlier) and the confinement
physics needed to give the proper quasiparticle currents on the
other, is in our opinion one of the most profound
questions facing the field today.

To summarize, the current status of the gauge theory approach to the
$t$-$J$ model is the following.  Within mean field
theory, a certain saddle-point solution may be regarded as the
``mother state'' at some intermediate energy scale.  The most
promising candidate for this ``mother state'' appears to be the
staggered flux state, which is the saddle point for the SU(2)
formulation of the $t$-$J$ model.  This saddle point is an unstable
fixed point due to gauge fluctuations.  It flows to the
N\'{e}el state at or near half filling and to the superconducting
ground state at finite doping.  This picture is in
accordance with a recent insightful comment by Anderson,\cite{15}
except that we provide a concrete description of the unstable
fixed point. As discussed above, the issues raised by this picture
are profound and probably difficult to treat
analytically.  In the past few years, we have focused on trying to
substantiate this point of view by numerical methods
and by predicting new experiments.  On the numerical front, the study
of projected wavefunctions have yielded remarkable
insights.  Ivanov {\it et al}.\cite{16} reported a finding which is
surprising except from the SU(2) gauge theory point of
view. They discovered that upon Gutzwiller projection of a BCS
$d$-wave state, the current-current correlation shows a
staggered pattern, as expected for the staggered flux state.
    Thus there is strong numerical
evidence that the staggered flux state and the
$d$-wave superconducting state are intimately related.  On the
experimental front, predictions have been made that the staggered
flux state may be stabilized in the vortex core and ways
to measure the effect of the orbital currents have been proposed.\cite{17}
In this paper we continue work in this direction and ask the question:
{\it is the superconducting ground state that emerges out of the gauge theory
completely conventional, or are there detectable consequences of its
unconventional origin}?

The short answer to this question is that gauge theory predicts new
collective modes in the superconducting ground state
which have experimental consequences.  Unlike conventional BCS
theory, where the only order parameter is the complex pairing
order parameter $\Delta$, here the hopping matrix element $\chi$ also
functions as an order parameter.  Thus it is natural to
expect new collective modes.  This problem is formulated generally in
Sections II and III.  In Section IV we show how the
gauge theory allows us to predict the low lying collective modes.
The new modes are the $\theta$ mode and the transverse and
longitudinal $\phi$ modes.  The $\theta$ mode generates staggered
orbital current fluctuations and is related to the current
fluctuations found in the projected wavefunctions.  However, while
the projected wavefunctions give only equal time
correlation, here we obtain the full dynamical spectral function.
The $\phi$ gauge modes are new excitations related to {\it
amplitude} fluctuations of $\chi$ and $\Delta$.  In Sections V and VI
we present numerical results of the spectral functions
and describe experiments which may couple to them.

The approximation employed in this paper is to treat the system in
the Higgs phase by adopting the radial gauge for the
bosons.  Gauge fluctuations are treated at the Gaussian level and
nonanalytic corrections, such as instantons which may lead to
confinement, are neglected. This can be justfied in terms of the 
$1/N$-expansion with $N$ being the number of fermion species.
In the present case, $N=2$ and one criterion for the validity of 
this Gaussian approximation is that the magnitude of the order
parameter fluctuation is less than that of the mean field value.
As is evident from the discussion in II and III, the magnitude of the 
fluctuation diverges as $x \to 0$ because the action for the SU(2) 
local gauge transformation vanishes in this limit. Hence 
the Gaussian approximation breaks down near $x=0$, 
where the confinement physics is of vital importance
because the nonperturbative configurations such as instantons
contribute appreciably. Below we will show mainly the numerical 
results for  $x=0.1$. We have estimted numerically the 
magnitude of the fluctuation of the order parameters, e.g.,
$< (\delta \chi_x)^2>$ at zero temperature, and found that 
it is of the order of unity at $x=0.1$. 
Therefore the gaussian approximation is  
marginaly justified in this
case. As another test, recently Honerkamp and Lee \cite{honerkamp}
have computed the shift in $\Delta$  using a free energy which includes
the gaussian fluctuaions of the $\theta$ mode. They found a reduction 
to 55 $\%$ of the mean field saddle point value
for x=.06 but for x=.1 the reduction is only 80 $\%$. Thus the effect of the
gaussian fluctuation is relatively under control for increasing doping.
Furthermore, since the collective modes appear at
fairly high energy $(\sim J)$ for experimentally relevant doping $(x
\approx 0.1)$, it is possible that this approximation is valid
at this energy scale, but the ultimate test must come from experiment.

\section{U(1) and SU(2) Formulations of the $t$-$J$ Model}
We begin with the familiar U(1) formulation of the $t$-$J$
model\cite{18}

\be
H= \sum_{<i,j>} J\left(
{\bbox{S}}_i\cdot {\bbox{S}}_j-
{1 \over 4} n_i n_j \right)
-t_{ij} c_{i\sigma}^\dagger c_{j\sigma}.
\label{t-J}
\en

The constraint of no double occupation is enforced by writing

\be
c_{i\sigma}^\dagger = f_{i\sigma}^\dagger b_i
\en
and imposing the condition
$
\sum_\sigma f_{i\sigma}^\dagger f_{i\sigma} + b^\dagger_i b_i = 1
$,
which in turn is enforced with a Lagrangian multiplier $\lambda_i$.
The Heisenberg exchange
term is written in terms of $f_{i\sigma}$\cite{19}

\be
{\bbox{S}}_i\cdot {\bbox{S}}_j &=& -{1\over 4} f_{i\sigma}^\dagger f_{j\sigma}
f_{j\beta}^\dagger f_{i\beta}  \nonumber  \\
&-& {1\over 4} \left(
f_{i\up}^\dagger f_{j\down}^\dagger - f_{i\down}^\dagger f_{j\up}^\dagger
\right) \left(
f_{j\down} f_{i\up} - f_{j\up} f_{i\down}
\right) \\
&+& {1\over 4} \left( f_{i\alpha}^\dagger f_{i\alpha}  \right) .\nonumber
\en
We write

\be
n_in_j = (1 - b_i^\dagger b_i ) (1 - b_j^\dagger b_j )  .
\en
Then ${\bbox{S}}_i\cdot {\bbox{S}}_j - {1\over 4}n_in_j$ can be
written in terms of the first two
terms of Eq. (3)  plus quadratic terms, provided we ignore the nearest-neighbor
hole-hole interaction ${1\over 4}b_i^\dagger b_ib_j^\dagger b_j $.
We then decouple the
exchange term in both the particle-hole and particle-particle channels via the
Hubbard-Stratonovich transformation.  By introducing the SU(2)
doublets\cite{20,21}

\be
\Phi_{i\up} =
\pmatrix{f_{i\up} \cr
f_{i\down}^\dagger} \,\, , \,\,\,\,
\Phi_{i\down} =
\pmatrix{f_{i\down} \cr
-f_{i\up}^\dagger} \,\, ,
\en
the partition function is written in the compact form

\be
Z = \int{D\Phi D\Phi^\dagger Db D\lambda DU} \exp \left(
-\int^\beta _0 d\tau L_1
\right)
\en
where

\be
L_1 &=& {\tilde{J}\over 2} \sum_{<ij>}Tr [U_{ij}^\dagger U_{ij}] +
{\tilde{J}\over 2} \sum_{<ij>,\sigma} \left (
\Phi_{i\sigma}^\dagger U_{ij} \Phi_{j\sigma} + c.c.
\right) \nonumber \\
&+&\sum_{i\sigma}f_{i\sigma}^\dagger (\partial_\tau - i\lambda_i)
f_{i\sigma} \nonumber \\
&+&\sum_i b_i^\ast(\partial_\tau - i\lambda_i + \mu ) b_i \nonumber \\
&-& \sum_{ij}{t}_{ij}b_ib_j^\ast f_{i\sigma}^\dagger f_{j\sigma} ,
\en
\be
U_{ij}=
\pmatrix{-\chi_{ij}^\ast&\Delta_{ij}\cr
\Delta_{ij}^\ast&\chi_{ij}}
\en
with $\chi_{ij}$ representing fermion hopping and $\Delta_{ij}$
representing fermion pairing.
In Eq. (7) $\tilde{J}_{ij}=J/4$ but in the
literature $\tilde{J}$ has
sometimes been taken to be $3J/8$.  The latter has the advantage that
the mean field equation
reproduces that which is obtained by the Feynman variational
principle,\cite{22} but these differences
are well within the uncertainties of the mean field theory.
( In the mean field theory, 
$\chi_{ij} = \sum_\sigma < f^\dagger_{i \sigma} f_{j \sigma} >$
and $\Delta_{ij} = < f_{ i \uparrow} f_{j \downarrow}
- f_{ i \downarrow} f_{j \uparrow}> $.)

Affleck {\it et al.}\cite{20} pointed out that the $t$-$J$ model at
half-filling obeys an exact SU(2) symmetry in the functional 
integral formulation.  The SU(2) doublet in Eq. (5) expresses the
physical idea that a physical up-spin can be represented by
the presence of an up-spin fermion, or the absence
of a down-spin fermion, once the constraint is imposed.
Wen and Lee\cite{23} proposed a formulation
which obeys the SU(2) symmetry even away from half-filling.
The SU(2) and the original U(1)
formulation are equally exact, but once approximations are
introduced, the SU(2) formulation
has the advantage that the zero doping limit $x \rightarrow 0$ can be
smoothly taken.  We
shall see an example of this in the collective mode spectrum
described below.  In the SU(2)
formulation a doublet of bosons is introduced

\be
h_i =
\pmatrix{b_{1i} \cr
b_{2i}} \,\,\, .
\en
The physical Hilbert space is the SU(2) singlet subspace. The
electron operator is an SU(2)
singlet formed out of the fermion and boson doublets

\be
c_{i\sigma} = {1\over \sqrt{2}} h_i^\dagger \Phi_{i\sigma}
\en
and three Lagrangian multipliers $a_{0i}^\ell$, $\ell = 1,2,3$ are
needed to project to the SU(2)
singlet subspace and impose the constraints

\be
{1\over 2} \Phi_{i\sigma}^\dagger {\bbox{\tau}}\Phi_{i\sigma} +
h_i^\dagger {\bbox{\tau}} h_i = 0
\,\,\, .
\en

Now the partition function $Z$ is given by  
\be
Z = \int{D\Phi D\Phi^\dagger Dh  Da^1_0 Da^2_0 Da^3_0 DU} 
\exp \left(-\int^\beta _0 d\tau L_2
\right)
\nonumber
\en
with the Lagrangian taking the form

\be
L_2 &=& {\tilde{J} \over 2} \sum_{<ij>} Tr \left[ U_{ij}^\dagger
U_{ij} \right] +
{\tilde{J} \over 2} \sum_{<ij>,\sigma} \left( \Phi_{i\sigma}^\dagger
U_{ij} \Phi_{j\sigma} +
c.c. \right) \nonumber \\
&+& {1\over 2} \sum_{i\sigma} \Phi_{i\sigma}^\dagger \left(
\partial_\tau - ia_{oi}^\ell \tau^\ell \right) \Phi_{i\sigma} \nonumber \\
&+& \sum_i h_i^\dagger \left(
\partial_\tau - ia_{oi}^\ell \tau^\ell + \mu \right) h_i \nonumber \\
&-&{1\over 2}\sum_{ij,\sigma} t_{ij} 
\Phi_{i\sigma}^\dagger h_i h_j^\dagger \Phi_{j\sigma}
\,\,\, .
\en

As pointed out in Ref. (24), Eq. (12) is closely related to the
U(1) Lagrangian Eq. (7)
if we transform to the radial gauge, i.e, we write

\be
h_i = g_i
\pmatrix{b_{i} \cr
0}
\en
where $b_i$ is complex and $g_i$ is an SU(2) matrix parametrized by

\be
g_i=
\pmatrix{z_{i1} & -z_{i2}^\ast\cr
z_{i2} & z_{i1}^\ast}.
\en
where

\be
z_{i1}  = e^{i\alpha_i} e^{-i{\phi_i\over 2}} \cos {\theta_i \over 2}
\en
and

\be
z_{i2}  = e^{i\alpha_i} e^{i{\phi_i\over 2}} \sin {\theta_i \over 2} \,\,\, .
\en
The angle $\alpha_i$ in $z_{i1}$ and $z_{i2}$ is the overall phase
which is redundant and can
be absorbed in the phase of $b_i$.

An important feature of Eq. (12) is that $L_2$ is invariant under the
SU(2) gauge transformation

\be
\tilde{h}_i &=& g_i^\dagger h_i \\
\tilde{\Phi}_{i\sigma} &=& g_i^\dagger \Phi_{i\sigma} \\
\tilde{U}_{ij} &=& g_i^\dagger U_{ij} g_j
\en
and

\be
\tilde{a}_{0i}^\ell \tau^\ell = g_i^\dagger a_{0i}^\ell \tau^\ell g_i 
+ i g_i^\dagger \left(
\partial _\tau g_i \right) \,\,\, .
\en
Starting from Eq. (12) and making the above gauge transformation, the
partition function is
integrated over $b_i$ and $g_i$ instead of $h_i$ and the Lagrangian
takes the form

\be
L_2^\prime &=& {\tilde{J} \over 2} \sum_{<ij>} Tr \left(
U_{ij}^\dagger U_{ij} \right) +
{\tilde{J} \over 2} \sum_{<ij>,\sigma} \Phi_{i\sigma}^\dagger
U_{ij}\Phi_{j\sigma} + c.c.
\nonumber \\
&+& {1\over 2} \sum_{i,\sigma} \Phi_{i\sigma}^\dagger \left(
\partial_\tau - ia_{0i}^\ell \tau^\ell  \right) \Phi_{i\sigma}\nonumber \\
&+& \sum_i b_i^\ast \left(
\partial_\tau - ia_{0i}^3 + \mu \right)b_i \nonumber \\
&-&  \sum_{ij,\sigma} \tilde{t}_{ij} b_i^\ast b_j
f_{j\sigma}^\dagger f_{i\sigma}
\en
We have removed the tilde from $\tilde{U}_{ij}$,
$\tilde{\Phi}_{i\sigma}$, $\tilde{f}_{i\sigma}$, $\tilde{a}_0^\ell$
because
these are integration variables.  Note that
$g_i$ has disappeard from the actions and Eq. (21) is the same as the
U(1) Lagrangian  $L_1$, with the exception that
$t_{ij}$ is now replaced by $\tilde{t}_{ij} = t_{ij}/2$, $\lambda_i$
becomes $a_{oi}^3$ and, most
importantly, two additional integrals $a_{0i}^1$ and
$a_{0i}^2$ coupling to the fermions appear.   We note that in the
limit of zero doping, thanks to these additional gauge
fields, Eq. (21) manifestly invariant under SU(2) transformation,
whereas Eq.(7) is not.   This will have important
consequences when we consider mean field approximation and small
fluctuations, in  that Eq. (21) will have smooth $x
\rightarrow 0$ limit while Eq. (7) does not.

\section{Mean Field Theory and Collective Modes}

We now consider the mean field treatment of Eq. (21) and the
quadratic fluctuations about the mean field, which will yield
the collective modes.  We work in
the radial gauge, where $b_i$ is considered real without loss of
generality.    As
discussed after Eq. (16), the phase of $b_i$ and $\alpha_i$ are
redundant and one of them can be chosen as zero.
We will discuss the alternative choice later, but here we choose $b_i$
to be real.   The saddle point solutions are

\be
b_i &=& r_0 \nonumber \\
\chi_{ij} &=& \chi_0 \nonumber \\
\Delta_{i,i+\mu} &=& \Delta_0 \eta_\mu \,\, 
 \mbox{{\rm or}} \,\,
\Delta_{i,j} = \Delta_0(-1)^{i_y+j_y} \nonumber \\
ia_{0i}^\ell &=& (0, 0, \lambda_0 )
\en
where $\mu = \hat{x}$ or $\hat{y}$ and $\eta_x = 1, \eta_y = -1$
correspond to $d$-wave pairing of the fermions.  The saddle
point corresponds to a physical
$d$-wave superconductor, as the order parameter 
$\langle c_{i\up} c_{i+\mu\down} - c_{i\down} c_{i+\mu\up} 
\rangle = r_0^2 \Delta_0 \eta_\mu$ is nonzero.
The mean field fermionic action is

\be
L_0 = -\sum_{{\bbox{k}}}
\pmatrix{f_{{\bbox{k}}\up}^\dagger \cr
f_{-{\bbox{k}}\down}}
\pmatrix{i\omega_n - \xi_{\bbox{k}} &,& -\Delta_k \cr
-\Delta_k &, &  i\omega_n + \xi_{\bbox{k}} }
\pmatrix{
f_{{\bbox{k}}\up} \cr
f_{-{\bbox{k}}\down}^\dagger }
\en
where
$\chi_{\bbox{k}} = 2{\tilde{J}}\chi_0 (\cos k_x + \cos k_y)$,
$\Delta_{\bbox{k}} = 2{\tilde{J}}\Delta_0 (\cos k_x - \cos k_y)$,
$t_{\bbox{k}} = 2{\tilde{t}} r_0^2 (\cos k_x + \cos k_y)$ and
$\xi_{\bbox{k}} = -\chi_{\bbox{k}} - t_{\bbox{k}} - \lambda_0$.
We write the small expansion about these saddle points as

\be
b_i &=& r_0 (1 + \delta R_i) \nonumber \\
U_{ij} &=& U_{ij}^d + \delta U_{ij} \nonumber \\
a_{0i}^\ell &=& (\delta a_{0i}^1, \delta a_{0i}^2, -i\lambda_0 + \delta
a_{0i}^3)
\en
where the mean field $U^d$ describes $d$-wave pairing,

\be
U_{i,i+\mu}^d = -\chi_0 \tau^3 + \Delta_0 \eta_\mu \tau^1
\rule{.55in}{0in} \,\,\, .
\en
The fluctuation is expanded as

\be
\delta U_{i,i+\mu} = \sum_{a=o}^3 \delta U_\mu^a \tau^a  \rule{.85in}{0in}
\en
where $\tau^0 = I$.  Note $\delta U_x^a$ and $\delta U_y^a, a = 1
\,\,{\rm to} \,\, 3$ are real variables, while for $a=0$ are purely
imaginary, and together make up a total of
8 degrees of freedom.  These correspond to complex hopping
$\chi_{i,i+\mu}$ and pairing $\Delta_{i,i+\mu}$ in the $x$ and $y$
directions.

By setting the linear terms in the small expansion of the free energy
to zero, we obtain the standard saddle point equations
A.1 to A.3.  The second order deviation is described by a $12 \times
12$ matrix, where the variables are $\delta U_\mu^a$,
$\delta R$, and $\delta a_0^\ell$.  The details are given in the
appendix.  As it stands the matrix is not hermitian.  On the other hand,
if we consider $ia^\ell_0$ as variables, the matrix is hermitian for
$\omega_n = 0$ but has negative eigenvalues, i.e., 
it corresponds to the saddle point of the free
energy in the {\it unprojected} Hilbert space.  In order to obtain
positive eigenvalues, it is necessary to first integrate out
$\delta a_{0i}^\ell$ in order to project to the physical subspace.  Since
this is a Gaussian integration, we may equivalently consider $(\delta
U_\mu^a ,
\delta R
$) as the physical degrees of freedom, and solve for the
local $\delta a_{0i}^\ell$ for each configuration.  This is the idea
behind the $\sigma$-model approach  in ref. (24) where
large fluctuations in $\delta U_{ij}$ are considered.  The present
work should be considered the low-temperature limit of the
$\sigma$-model.

At this point we proceed numerically, evaluate the $12 \times 12$
matrix and integrate out the $\delta a_0^\ell$ fields.
The remaining quadratic form gives a $9 \times 9$ matrix with 9
eigenvalues.  As expected, there is a soft mode associated
with the phase of the pairing order parameter.  In addition, we find
a number of soft modes in the small $x$ limit.  Before
presenting the numerical results, we show how the SU(2) symmetry
allows us to predict and classify all the soft degrees of
freedom.

\section{SU(2) Classification of Soft  Modes}
In this section we make use of the SU(2) gauge symmetry to classify
the soft modes.  The basic idea is the  following.  As the
temperature is decreased, the SU(2) symmetry is broken via a series
of symmetry breaking at the mean field level.  For small
$x$, SU(2) is first broken down to U(1) at a temperature scale of
order $\tilde{J}/2$ to the staggered flux ($s$-flux) state.\cite{23}
At a lower temperature of order $xt$, the bosons condense and the
gauge symmetry is broken completely.  Of course, a
local gauge symmetry cannot be broken, but the mean field description
is still a useful starting point to describe the
low-lying collective excitations, which are physical.  A familiar
example is the pairing order parameter of superconductivity,
which breaks the local U(1) gauge symmetry associated with  the E\& M
gauge field at the mean  field level. While strictly
speaking, this order parameter is not gauge invariant, it is a useful
starting point which leads to the correct description of the gauge field
via the Anderson-Higgs mechanism.

It is useful to distinguish
between two kinds of symmetry breaking as $x \rightarrow 0$.  First,
at $x = 0$ the mean field solution is the $\pi$-flux
state, i.e., $\chi_0 = \Delta_0$.  This state has full SU(2)
symmetry, which is broken down to U(1) in the $s$-flux state,
where $\chi_0 \neq \Delta_0$.  We shall refer to the remaining
symmetry as the residual U(1) symmetry.  As $x$ becomes
nonzero, $\Delta_0/\chi_0$ deviates from unity rather rapidly and we
shall focus our attention on the zero modes due to the
residual U(1) symmetry.  Secondly, the bosonic degrees of freedom
appear at $x \neq 0$. Boson condensation breaks the residual
U(1) completely below an energy scale of order $xt$.  The zero modes
then acquire a finite energy gap which is the subject of
our analysis.

Starting from the $d$-wave superconductor mean field solution
described in the last section, we expect the soft modes to
involve small fluctuations of the boson about the radial gauge which
can be parametrized by the SU(2) matrix $g_i$ such that
$h_i = g_i \pmatrix {r_0 \cr 0}$. In addition, we include phase
fluctuation of the
$U_{ij}$ matrix which we paramaterize by

\be
U_{ij} = U_{ij}^d e^{i {\bbox{a}}_{ij} \cdot {\bbox{\tau}}}
\en
where $a_{ij}^\ell,$ $\ell = 1,2,3$ are three gauge fields
living on spatial links.  Since SU(2) has been broken down
to U(1), only one out of three  gauge fields remain soft in the
$s$-flux state.\cite{25,26} To visualize this, it is convenient to make a
gauge transformation using Eqs. (17, 18, 19) to the $s$-flux order parameter

\be
U_{ij}^{SF} = w_i^\dagger U_{ij}^d w_j
\en
where

\be
w_j = \exp \left[
i(-1)^{j_x+j_y} {\pi\over 4} \tau^1 \right]
\en
and

\be
U_{ij}^{SF} &=& -\chi_0 \tau^3 - i\Delta_o (-1) ^{i_x+j_y} \nonumber \\
&=& -A \tau^3 \exp \left(
i(-1)^{i_x+j_y} \Phi_o \tau^3 \right)
\en
where $\chi_0 = A \cos\Phi_0$ and $\Delta_0 = A\sin\Phi_0$.  Equation
(30) represents fermion hopping with a complex matrix
element such that a flux $4 \Phi_0$ threads the lattice plaquettes in a
staggered manner.  At the same time the boson is
transformed to

\be
h_i^{SF} = w_i^\dagger
\pmatrix { r_0 \cr
0}
= { {r_0} \over {\sqrt{2} } }
\pmatrix { 1 \cr
-i(-1)^{i_x+i_y}}
\en
and the mean field ${\bbox{a}}_0$ becomes  ${\bbox{a}}_0^{SF} =
(0, i \lambda_0 (-1)^{i_x+i_y}, 0)$.  We note that $U_{ij}^{SF}$ describes a
semiconductor band with nodes at $(\pi/2,\pi/2)$.   If $h_i$ were to remain as
$
h_i = r_0
\pmatrix {1 \cr 0}
$
and
$a_0^3\neq 0$
,
we would have described an $s$-flux state with small fermion pockets.
Instead, $h_i$ and
     ${\bbox{a}}_0^{SF}$ are rotated such that ${\bbox{a}}_0^{SF}$
couples to pair fields
$f_{i\up}^\dagger f_{i\down}^\dagger$ and the resulting state is
gauge equivalent to the $d$-wave superconductor that we
started out with.  The advantage of the $s$-flux representation is
that $U_{ij}^{SF}$ is proportional to $\tau_3$ and is
invariant under $\tau_3$ rotation.  Thus the residual U(1) symmetry
is apparent.  We expect the soft modes to be described by

\be
{\tilde{U}}_{ij}^{SF} = U_{ij}^{SF} e ^{ia_{ij}^3\tau^3}
\en
and

\be
h_i^\prime = g_i^\prime
\pmatrix {r_0 \cr
0}
\en
where $h_i^\prime$ is close to  $h_i^{SF}$, i.e.,
$g_i^\prime$ is parametrized by Eq. (14) with $\theta$ close to
${\pi\over 2}$ and $\phi$ close to $(-1)^{i_x+i_y}{\pi\over 2}$.  In
Eq. (32) we have ignored the $a_{ij}^1$ and $a_{ij}^2$
gauge fields as they have been pushed to finite frequencies by the
Anderson-Higgs mechanism.

In order to visualize the different gauge choices, it is useful to
introduce  the local quantization axis

\be
{\bbox{I}}_i = z_i^\dagger {\bbox{\tau}} z_i =
(\sin\theta_i\cos\phi_i , \sin\theta_i\sin\phi_i , \cos \theta_i)
\,\,\, .
\en
Note that ${\bbox{I}}_i$ is independent of the overall phase $\alpha_i$.
In the $s$-flux representation the quantization axis has been rotated with
$w_i$ given by Eq.(29) to point along the $\pm y$-axis in a
staggered fashion.  Small fluctuations correspond to $\delta\theta$
deviation from the equator and $\delta\phi$ in the
azimuthal angle.  This is illustrated in Fig.1.

\begin{figure}[btp]
\centerline{
\psfig{figure=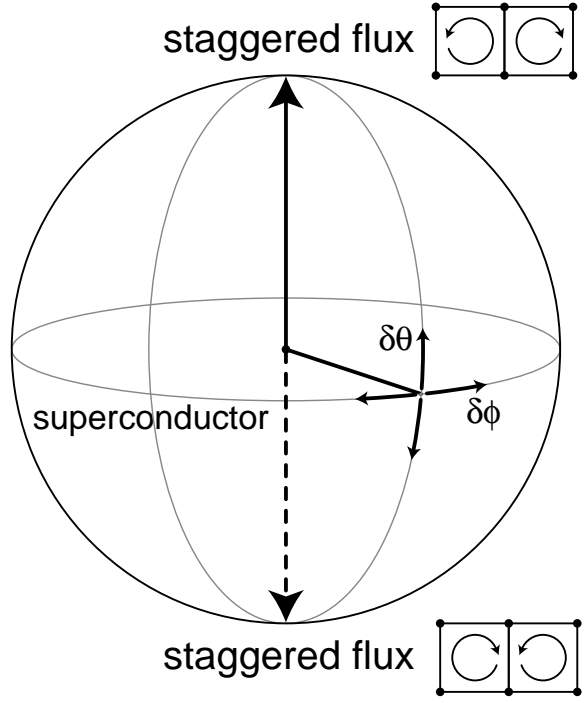,width=3.in}
}
\vspace{0.5cm}
\caption{The quantization axis $\bbox{I}$ in the SU(2) gauge theory.
The north and
south poles correspond to the staggered flux phases with shifted
orbital current
patterns.  All points on the equators are equivalent and correspond
to the $d$-wave
superconductor.  In the superconducting state one particular
direction is chosen on the
equator.  There are two important collective modes.  The $\theta$
modes correspond to
fluctuations in the polar angle $\delta\theta$ and the $\phi$ gauge
mode to a spatially
varying fluctuation in $\delta\phi$.}
\end{figure}

It is useful to rotate the configuration specified by Eqs.(32) and
(33) back to the radial gauge.  We obtain

\be
{\tilde{U}}_{ij}^d = g_i^{\prime\dagger}U_{ij}^{SF}
e^{ia_{ij}^3\tau^3} g_j^\prime \,\,\, .
\en
The advantage of the radial gauge is two-fold.  The electron operator
$c_{i\sigma}=r_0 f_{i\sigma}$ and we can consider
${\tilde{U}}_{ij}^d$ as the effective Hamiltonian for electron
quasiparticles.  At the same time, we can  now make contact
with the fluctuation of the $U_{ij}$ matrix in the last section and
interpret the numerical results.

Equation (35) can be explicitly evaluated for arbitrary
$\theta$, $\phi$ and $\alpha$ and $a^3_{ij}$, resulting in

\be
\tilde{U}_{ij}^d =
\pmatrix{
\hat{U}_{11}^d e^{-i(\alpha_i-\alpha_j)} &
\hat{U}_{12}^d e^{-i(\alpha_i+\alpha_j)} \cr
\hat{U}_{21}^d e^{i(\alpha_i+\alpha_j)} &
\hat{U}_{22}^d e^{i(\alpha_i-\alpha_j)}
}.
\en

The overall phase $\alpha_i$ enters the effective hopping
$\tilde{U}_{22}^d$ and
effective pairing $\tilde{U}_{12}^d$ in the expected way, and
$\hat{U}^d_{ij}$ is
the $\alpha_i=0$ limit given by ref. [27].

\be
{\hat{U}}_{ij}^d &=& -{\tilde{\chi}}_{ij} \left[
\tau^3 \cos {\theta_i-\theta_j \over 2} + (-1)^{i_x+i_y} \tau^2 \sin
{\theta_i-\theta_j \over 2} \right] \nonumber \\
&& - {\tilde{\Delta}}_{ij} \left[ \rule{0in}{.15in}
i(-1)^{i_x+i_y} \cos {\theta_i+\theta_j \over 2} -\tau^1 \sin
{\theta_i+\theta_j \over 2} \right]
\en
where

\be
{\tilde{\chi}}_{ij} &=& A \cos \tilde{\Phi}_{ij} \nonumber \\
{\tilde{\Delta}}_{ij} &=& A \eta_{j-i}  \sin\tilde{\Phi}_{ij} \\
{\tilde{\Phi}}_{ij} &=& \Phi_0 + (-1)^{i_x+j_y} v_{ij} \nonumber
\en
and

\be
v_{ij} = {\phi_i-\phi_j \over 2} - a_{ij}^3 \,\,\, .
\en
Note that the only dependence on $\phi_i$, $\phi_j$ is via the gauge
invariant combination $v_{ij}$, which has the
interpretation of the gauge current.  Furthermore, for
$\theta_i=\theta_j = {\pi\over 2}$, we see from Eqs.(36) and (37) that
${\tilde{\chi}}_{ij}e^{i(\alpha_i-\alpha_j)}$ and ${\tilde{\Delta}}_{ij}
e^{-i(\alpha_i+\alpha_j)}$ play the role of the effective hopping and pairing
parameters.  Thus fluctuations in $\phi_i$ leading to
nonzero $v_{ij}$ means a fluctuation in the {\it amplitude} of
${\tilde{\chi}}_{ij}$ and ${\tilde{\Delta}}_{ij}$ in such a way
     that $|{\tilde{\chi}}_{ij}|^2 + |{\tilde{\Delta}}_{ij}|^2 = A^2$ is
fixed.  This is shown in Fig.2.

\begin{figure}[btp]
\centerline{
\psfig{figure=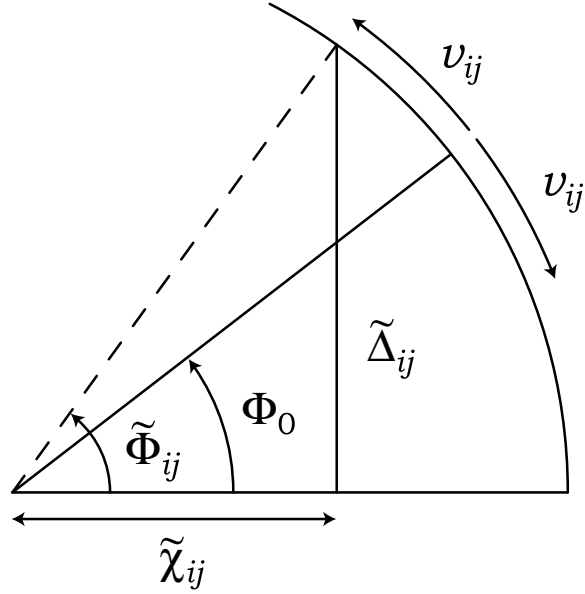,width=3.in}
}
\vspace{0.5cm}
\caption{Geometrical interpretation of the fluctuation of
$v_{ij}$.  The angle $\tilde{\chi}_{ij}$ [given by Eq.(38)]
is modulated around the flux $\Phi_0$ in a staggered
manner, in such a way that the hopping amplitude
$\tilde{\Phi}_{ij}$ and the pairing amplitude
$\tilde{\Delta}_{ij}$ are modulated, also in a staggered
manner.
}
\end{figure}

To make contact with the collective modes, we write $\theta_i =
{\pi\over 2} + \delta\theta_i$ and expand Eqs.(36,37) to first
order in $\delta\theta_i$, $v_{ij}$ and $\alpha_i$.

\be
{\tilde{U}}_{ij}^d - U_{ij}^d &=& -\chi_0 \tau^2 {(-1)^{i_x+i_y}\over
2} (\delta\theta_i - \delta\theta_j) + i\Delta_0 \eta_\mu
{(-1)^{i_x+i_y}\over 2} (\delta\theta_i + \delta\theta_j) \nonumber \\
&& + \left( \chi_0 \tau^1 + \Delta_0 \eta_\mu \tau^3 \right)
{(-1)^{i_x+i_y}\over 2} v_{ij} \nonumber \\
&& + i { 1 \over 2} \chi_0 (\alpha_i  - \alpha_j) \nonumber \\
&& + { 1 \over 2} \Delta_0 \eta_\mu \tau^2 (\alpha_i  + \alpha_j)
\en
where $j = i + \mu$, $\mu = \hat{x}, \hat{y}$.

Equation (40) is a main result of this section, as it allows us to
interpret the collective modes when compared with Eq.(26).
It predicts the location in momentum space of the soft collective
modes and gives the eigenvectors. We classify the modes as
follows:

\begin{eqnarray}
\mbox{\rm {uniform}} \rule{.05in}{0in} \alpha : \alpha_i &\approx& \alpha_u ,
\nonumber \\
\mbox{\rm{staggered}} \rule{.05in}{0in} \alpha : \alpha_i &\approx&
\alpha_s(-1)^{i_x+i_y} ,  \nonumber \\
\mbox{\rm{uniform}} \rule{.05in}{0in} \theta : \theta_i &\approx& \theta_u ,
\nonumber \\
\mbox{\rm{staggered}} \rule{.05in}{0in} \theta : \theta_i &\approx&
\theta_s(-1)^{i_x+i_y} ,
\end{eqnarray}

We consider $\alpha_u$, $\alpha_s$, $\theta_u$, $\theta_s$ and
$v_{i,i+\mu}(\mu=x,y)$ as six slowly varying variables and the
first four variables are defined as follows

\begin{eqnarray}
{1\over 2} \left( \alpha_i + \alpha_j \right) &=& \alpha_u \nonumber \\
{1\over 2} \left( \alpha_i - \alpha_j \right) &=&
(-1)^{i_x+i_y}\alpha_s  \nonumber \\
{1\over 2} \left( \theta_i + \theta_j \right) &=& \theta_u \nonumber \\
{1\over 2} \left( \theta_i - \theta_j \right) &=&(-1)^{i_x+i_y}\theta_s
\end{eqnarray}
     Substitution into Eq.(40) and comparison with Eq.(26) show that at
$\bbox{q}=0$, we can identify the following collective
modes and their eigenvectors.

\begin{equation}
\begin{array}{lcc}
\mbox{\rm {Goldstone mode:}}      &\alpha_u &, \rule{.15in}{0in}U_x^2
- U_y^2 \\
\mbox{\rm{internal phase mode:}} &\theta_s   &, \rule{.15in}{0in} U_x^2 + U_y^2
\end{array}
\end{equation}
At $\bbox{q}=(\pi,\pi)$ we pick out the coefficients of
$(-1)^{i_x+i_y}$ and identify the following modes and
corresponding eigenvectors.

\begin{equation}
\begin{array}{lllc}
\theta \,\,\, \mbox{\rm {mode:}}          &\theta_u        &,
\rule{.15in}{0in}U_x^0 - U_y^0  &   \\
\phi \,\,\, \mbox{\rm{gauge\,\,mode:}}         & v_{i,i+\mu}     &,
\rule{.15in}{0in} \chi_0 U_x^1 + \Delta_0 U_x^3
&, \rule{.15in}{0in} \chi_0 U_y^1 - \Delta_0 U_y^3  \\
\chi \,\,\, \mbox{\rm {mode:}}            &\alpha_s        &,
\rule{.15in}{0in}U_x^0 + U_y^0  &
\end{array}
\end{equation}
The notation $aU_x^\alpha + bU_y^\beta$ means that 
each component of the eigenvector is 
$\left(\delta U_x^\alpha = a, \delta U_y^\beta = b \right)$, etc.in Eq.(26) 
The nature of the collective modes are readily identified from their
eigenvectors.  Two of these modes were known before.  The
Goldstone mode $\alpha_u$ is the standard one associated with the
phase $\alpha_u$ of the superconducting order parameter
$\Delta_\mu e^{-i2\alpha_u}$.  The $\theta_s$ mode corresponds to the
out-of-phase oscillation of the phase of the
superconducting order parameter in the $x$ and $y$ directions,
$|\Delta_x|e^{-i2\varphi_x}$ and $|\Delta_y|e^{-i2\varphi_y}$, such
that $\theta_s = {1\over 2}(\varphi_x - \varphi_y)$.  This is a
property of any $d$-wave superconductor and is labelled the
internal phase mode.

The $\alpha_s$ mode corresponds to the fluctuation in the phase of
$\chi_{ij}$.  In BCS theory the hopping term is fixed and
not allowed to fluctuate.  So this is a new degree of freedom special
to the gauge theory.  The slowly varying phase of
$\chi_{ij}$ plays the role of the spatial component of the gauge
field in the U(1) gauge theory.\cite{18} This appears as a collective mode at
$\bbox{q} = (\pi,\pi)$.  We shall see that just as in the U(1) theory, the
$\alpha_{s}$
mode plays the crucial role in producing the correct answer for the superfluid
stiffness $\rho_s = x$.

The new modes that are of greatest interest to us in this paper are
the $\theta$ mode and the $\phi$ gauge modes.  These will
be discussed in greater detail later.

\begin{figure}[h]
\centerline{
\psfig{figure=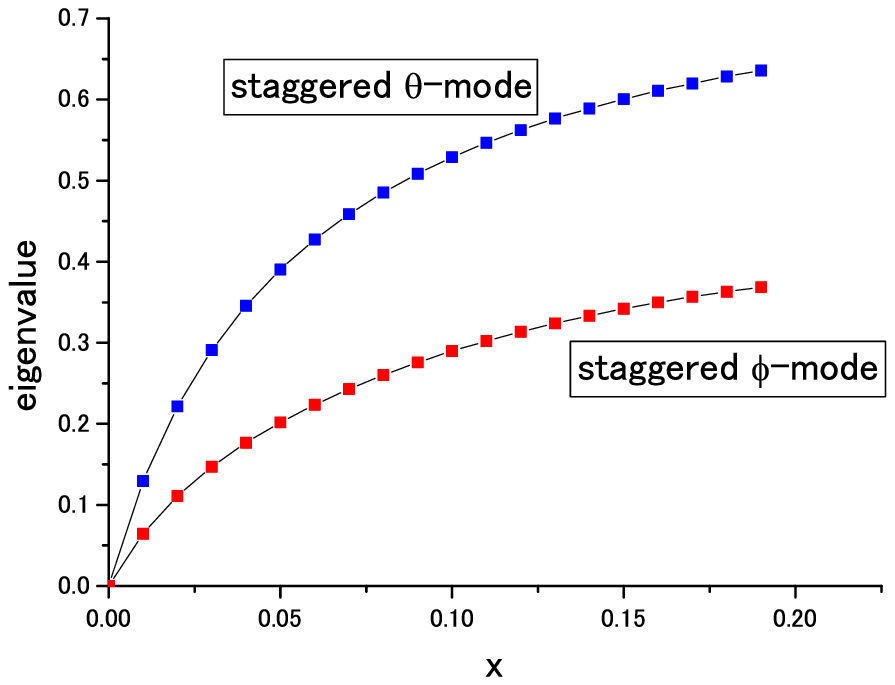,width=3in}
}
\caption{The eigenvalues at $\bbox{q} = (0,0)$ and $\omega=0$
of the $9 \times 9$ matrix
which describes fluctuations about the mean field as a function of doping
$x$. Shown are the eigenvalues which vanish as $x \rightarrow 0$. The
eigenvalues are given in units of ${\tilde{J}}$. In addition to the
Goldstone mode corresponding to superconducting phase fluctuations which has
zero eigenvalue for all $x$ (not shown), we find the staggered $\theta$
mode [called internal phase mode in Eq.(43)] and a continuation of the
transverse $\phi$-gauge mode from $\bbox{q} = (\pi,\pi)$.
   }
\end{figure}

Now we compare  our analysis with numerical results described in the
last section.  In Figs.3 and 4 we plot all the
eigenvalues of the $9\times 9$ matrix which vanishes at $x=0$ as a
function of $x$ at $\omega=0$ and at $\bbox{q}=0$ and
$\bbox{q}=(\pi,\pi)$, respectively.  They can all be identified with
our classification.  At $\bbox{q}=0$ the Goldstone mode
has zero eigenvalue for all $x$ as expected.  The internal phase mode
rises rapidly with increasing $x$.  In addition, we find
a mode with an eigenvector that corresponds to the
continuation of the transverse $\phi$ gauge mode we find at
$\bbox{q}=(\pi,\pi)$ to $\bbox{q}=0$.

\begin{figure}[tbp]
\centerline{
\psfig{figure=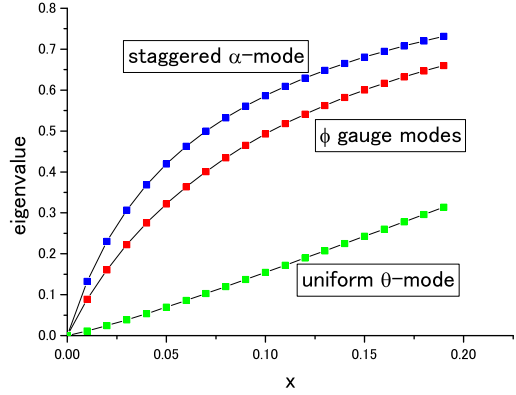,width=3in}
}
\caption{Same as Fig.3 but at $\bbox{q} = (\pi,\pi)$.  We find the
uniform $\theta$ mode, two degenerate $\phi$ modes, and a staggered
$\alpha$ mode [called the $\chi$ mode in Eq.(44)].
   }
\end{figure}

At $\bbox{q}=(\pi,\pi)$ we find the $\theta$ mode and the two $\phi$
gauge modes (transverse and longitudinal) which are degenerate.  In
addition we find the
$\chi$ mode.   The vanishing of the eigenvalues of all these modes at
$x=0$ is  a consequence of the SU(2) symmetry.  However,
as we discussed before, even if $|\chi| \neq |\Delta|$, as long as $x=0$
the $\theta$ mode still has zero eigenvalue and
the  $\phi$ gauge modes are
protected by the residual U(1) symmetry.  These modes are
gapped only by the boson condensation.

We should mention that in addition to these  soft modes, we found two
unstable eigenstates for small $x$.
If we use in Eq. (21)  ${\tilde t}/{\tilde J}=16/3$, i.e.,
${\tilde t}/J= 2$ and
$t^\prime = -0.5 t$, which generates a realistic Fermi surface,
the eigenvalue corresponding to $r_0$ fluctuation is negative for
$x < 0.084$.  Since $r_0$ corresponds to hole density, this
signals an instability to phase separation.  This instability is
easily suppressed by long range interaction and is largely
decoupled from the modes of interest.  A second instability occurs
for  $x < x_c$ near $\bbox{q}=(0,\pi)$ and $(\pi,0)$.  The
eigenvector corresponds to amplitude fluctuations in $|\chi|^2 +
|\Delta|^2$ which we identify as instability to columnar
dimer formation.  This instability is well known for $x=0$ and can be
suppressed by bi-quadratic terms.\cite{26}  For the parameters mentioned
above, we find $x_c = 0.087$.
Most of our detailed numerical results are for $x>x_c$.
Furthermore these two modes are mostly decoupled from the 
low lying modes in Eq.(41), and does not disturb their behavior 
in the limit $x \to 0$ discussed in the followings.

We next discuss in greater detail the $\theta$ and the $\phi$ gauge
modes. As seen in Fig.1, the $\theta$ mode is a fluctuation of $\bbox{I}$
towards the north and south poles,  and describes the admixture of
the $s$-flux phase.  From Eq.(37) we see that
$\theta_u$ generates a staggered imaginary part to the hopping matrix
elements which will produce staggered orbital currents.
Thus the physical manifestation of the $\theta$ mode is staggered
orbital current fluctuations.  The softness of the $\theta$
mode means that it is readily excited by thermal or quantum
fluctuations.  This is clearly related to the strong staggered
orbital current fluctuations found in the projected $d$-wave
superconductor wavefunction.\cite{16}  The energy cost  is low because the
$s$-flux  and the $d$-wave superconductor are almost degenerate in
energy.  The energy difference arises only because
$\bbox{a}_0$ is different in the two states.
At the mean field level, the energy difference comes from the Fermi
pockets and is proportional to $x^2$.  However, we find that after
integrating out $\bbox{a}_0$, the energy cost increases, apparently due to the
enforcement of the constraint and the eigenvalue shown in Fig.3 is linear
in $x$.
We have also computed the
eigenvalues for finite $\omega$. It should be
noted that the $9 \times 9$ matrix is not Hermitian for finite 
$\omega_m$(Matsubara frequency).
Then we make the analytic continuation as $ i \omega_m \to \omega + i \delta$
with infinitestimal $\delta>0$. This $\delta$ can be neglected outside of
the particle-hole continuum, and in this case the $9 \times 9$ matrix is
Hermitian and its eigenvalues are real. We will present below the results
for $(\bbox{q},\omega)$ outside of the particle-hole continuum.
We found that the eigenvalue can
be fitted by
$ ax - {b\over x} \omega^2 - {c \over x} {\bf k}^2 $
for small $\omega$ and $\bbox{k} = \bbox{q} -(\pi,\pi)$.
The inverse $x$ dependence of the
coefficient might be surprising but the $\omega^2$ and $\bf k$ region
shrinks as $x$ becomes small, so that the $x=0$ limit is smoothly attained.
Furthermore, the negative coefficient of $\bbox{k}^2$ indicates that the
eigenvalue has a local maximum at $\bbox{q} = (\pi,\pi)$.
  A plausible dependence is  $\sqrt{ (ax)^2 - b' \omega^2 - c'
{\bf k}^2} $, which is reduced to the above form in the limit 
$|\omega|,| {\bf k}|<<x$.   Back to the Matsubara frequency, the suggested
effective action for the $\theta$ mode is
\be
S_\theta = \sum_{\bbox{k},\omega_m}
\sqrt{ (ax)^2 + b' \omega_m^2 - c' {\bf k}^2 } |\theta (\bbox{k},\omega_m) |^2
\en
in the region $|\omega_m|, | {\bf k}|<<x$.
The small energy gap $ax$ leads
to a strong spectral weight.  This is confirmed by direct numerical
calculations in the next section.
Here it is interesting to compare these results with the SU(2) formalism
with those in the U(1) where only $a_0^3$ is integrated over. The
suggested action in this case is
\be
S^{U(1)}_\theta = \sum_{\bbox{k},\omega_m}
\left[ (ax)^2 + b^{\prime\prime} \omega_m^2 + 
c^{\prime\prime} x {\bf k}^2  \right]
|\theta (\bbox{k},\omega_m) |^2.
\en
Therefore it is evident that the SU(2) symmetry leads to quite different
$x$-dependence in the limit $x \to 0$.

To investigate the $\phi$ gauge modes we first discuss $x=0$.  The
mean field solution is the $\pi$-flux phase where the
fermions obey the Dirac spectra with nodes at $(\pm{\pi\over 2},
\pm{\pi\over 2})$.  After integrating out the fermions, it is
known that the effective gauge field action is purely transverse and 
given by\cite{28}

\be
S_\phi = \sum_{\bbox{k},\omega_m} \alpha_0 \sqrt{\bbox{k}^2 + \omega_m^2}
\left(
\delta_{\mu \nu} - {\bbox{k}_\mu \bbox{k}_\nu \over \bbox{k}^2} \right)
a_\mu(\bbox{k},\omega_m)a_\nu(\bbox{k}, - \omega_m)
\en
where ${\bbox{a}}_\mu$ is the continuum version of $a_{i,i+\mu}$ and
${\bbox{k}}$ is measured relative to $(\pi,\pi)$.  We confirm this by
computing the eigenvalues for
$x=0$ at finite $\bbox{q} = (1.03
\pi,\pi)$.  The transverse mode behaves as
$\sqrt{k^2+\omega_m^2}$ as expected while the longitudinal mode is
exactly zero for all $\omega_m$.   This is because the
longitudinal mode is pure gauge
$(\bbox{a}_\mu = \bbox{\nabla}_\mu \phi)$ and is not a real degree of
freedom.  In constrast, if we worked with the U(1)
formulation [Eq.(7)], we find that the longitudinal mode has
eigenvalue close to $|\omega|$.  Thus the addition of the $a_0^1$
and $a_0^2$ in the SU(2) formulation is crucial.  Otherwise we would
have gotten a  spurious collective mode.  This is a
dramatic illustration of the advantage of the SU(2) formulation, if
one is interested in obtaining meaningful results which are
smoothly connected to the undoped case.  For finite $x$ we see from
Fig.3 that the eigenvalue increases linearly with $x$ for
small $x$.  A reasonable approximate for the  transverse mode is
$\left[ \mbox{\rm for} \,\,\, \bbox{k} = \bbox{q} - (\pi,\pi)
\right]$

\be
S^{{\rm transverse}}_\phi = \sum_{\bbox{k},\omega_m}
\left(  \alpha_0\sqrt{(a_{\varphi}x)^2 + \bbox{k}^2 + \omega_m^2}
\rule{.05in}{0in}\right)
\left(  \delta_{\mu\nu} - {\bbox{k}_\mu  \bbox{k}_\nu \over \bbox{k}^2} \right)
v_\mu(\bbox{k},\omega_m) v_\nu(\bbox{k},\omega_m)
\en
where

\be
\bbox{v}_\mu = {1\over 2} \bbox{\nabla}_\mu \phi - \bbox{a}_\mu
\en
is the continuum limit of $v_{i,i+\mu}$ and $\alpha_0$ parameterizes
the spectral weight.  Thus we expect the transverse mode
to show an edge singularity starting at
$\omega=\sqrt{(a_\phi x)^2 + \bbox{k}^2 }$. In Fig.5 we show the
eigenvalues of the $\phi$ gauge modes as function of
$(q_x-\pi)/\pi$ at $q_y = \pi$ and $\omega_m=0$, which 
shows the expected dependence on $q_x-\pi$ from eq.(48) for
transverse mode. 

\begin{figure}[tbp]
\centerline{
\psfig{figure=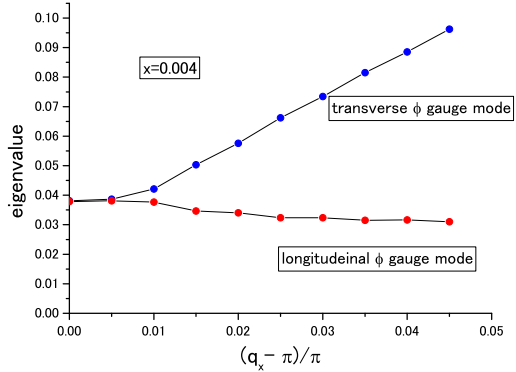,width=3in}
}
\caption{The eigenvalue (in units of ${\tilde{J}}$) of the transverse and 
longitudinal $\phi$ gauge mode as a function of 
$k_x/\pi  = (q_x - \pi)/\pi$ with $q_y = \pi$ and $\omega_m=0$ for $x =  
0.004$. 
They show the expected behavior for small $x$ and 
$\bbox{q}$ near $(\pi,\pi)$ because when $x=0$ and $\omega_m=0$, the 
transverse mode should be linear in $|\bbox{k}|$ and the longitudinal mode 
should be zero for all $\bbox{q}$ and $\omega_m$. 
    } 
\end{figure}

At finite $x$ the longitudinal mode becomes a real degree of freedom
due to the breaking of the residual U(1) symmetry.  At
$\omega_m=0$ and $\bbox{q}=(\pi,\pi)$ the eigenvalue is $a_\phi x$,
degenerate with the transverse mode.  For finite
$k_x = q_x - \pi$, the eigenvalue shows a slow decrease 
and then saturates. This is shown in Fig.5.  This is to be
expected as the eigenvalue is zero for all $\omega_m$
and $\bbox{q}$ for $x=0$. As for the spectral function, we just 
invert the $12 \times 12$ matrix numerically 
with $\omega_m \to \omega + i \delta$,
and we do not have to worry about the non-Hermitian nature of the 
matrix, which will be shown in the next
section.  It is worth noting that the longitudinal gauge mode is the
only mode with significant coupling to $\delta R$
and therefore to density fluctuations.

We next discuss the Goldstone mode.  We expect the eigenvalue to be
of the form $\rho_s (\bbox{\nabla\alpha})^2$, where
$\alpha$ is the continuum limit of a slowly varying $\alpha_u$.  We
indeed verify that upon diagonalizing the $9 \times 9$ matrix described in the
last section, a soft eigenvalue emerges which is linear in
$q^2$.  Furthermore the coefficient
$\rho_s$ is proportional to $x$.  This is to be expected in our
SU(2) formulation, as $x=0$ is an insulator.  It is interesting to
analyze how this result emerges.
The point is that the coefficient of $(\nabla\alpha_u)^2$ in the free energy
computed using Eqs.(21,40) is $\rho_f$, which corresponds to the
superfluid density
of a nearly half-filled conventional superconductor, and $\rho_f
\approx 1$.  How do we
obtain $\rho_s = x$ ?  In the numerics we find that the $\rho_sq^2$ mode is a
coupled mode between $\alpha_u$ and $\alpha_s$.  Recall that
$\alpha_s$ is related
to the phase of $\chi_{ij}$ and may therefore be identified with the U(1) gauge
field $\bbox{a}$ in the U(1) formulation.\cite{18}  In that
formulation the free energy in
the superconducting (Bose condensed) state takes the form

\be
F = \rho_f (\bbox{\nabla}\alpha_u - \bbox{a})^2 + \rho_b |\bbox{a}|^2
\en
where $\rho_b = x$ and $\rho_f$ is the fermion contribution to the superfluid
density, which is of order unity.  Upon minimizing $F$ with respect to
$\bbox{a}$, we arrive at the Ioffe-Larkin formula\cite{28} $F = 
\rho_s (\nabla\alpha_u)^2$ where

\be
\rho_s = {\rho_b\rho_f \over \rho_b + \rho_f} \,\,\, .
\en
The screening of $\nabla\alpha_u$ by the gauge field $\bbox{a}$ converts the
fermion response $\rho_f$ to the physical response $\rho_s \approx x$.  When we
express Eq.(50) in matrix form we see that the free energy has diagonal
contributions $\rho_f(\bbox{\nabla}\alpha_u)^2$ and $(\rho_f + \rho_b)
\bbox{a}^2$ and an off-diagonal term $-\rho_f \bbox{\nabla}\alpha_u \cdot
\bbox{a}$.  When we examine the $9 \times 9$ matrix, we find that for
small $q$,
$\alpha_u$ is only coupled to $\alpha_s$ and the form of the $2\times 2$
sub-matrix is just that given by the above discussion if we identify $\bbox{a}$
with $\alpha_s$.  Upon diagonalizing the $2 \times 2$ matrix, the
soft mode with
eigenvalue $\rho_s q^2$ emerges.  This simply confirms that our SU(2)
formulation contains the same U(1) gauge field and the same
screening mechanism is at work.  One consequence of
Ioffe-Larkin screening is that at finite temperature $\rho_s(T) =
\rho_s(0) - Cx^2 T$, i.e.,  the coefficient of the
linear $T$ term is proportional to $x^2$.  We have also verified
numerically that this is the case by computing $\rho_s$
at  finite $T$.  Experimentally, there is strong evidence that the
coefficient of the linear $T$ term is independent of
$x$.\cite{29}  The fact that $\rho_s$  is proportional to $x$ can be
seen more readily if we
associate the phase $\alpha_i$ with $b_i$ in Eq.(13)  instead of with
$z_i$ as we have done so far.
Then it is clear that for static $\alpha_i$, only the last term in
Eq.(12) depends on
$\alpha_i-\alpha_j$ and the free energy change must be
proportional to $r^2_0 \approx x$.  Furthermore, the
coupling to fermion excitations is proportional  to $x$, so
that the response to thermal excitations of quasiparticles
is proportional to $x^2$.
Thus Ioffe-Larkin and our  fluctuation theory are in
disagreement with experiment.  We believe this is an
indication that the fermions and bosons are confined to become
electrons in the superconducting state and that
confinement physics is beyond the Gaussian fluctuation considered here.

\section{Numerical Results for the Spectral Functions of the $\theta$
Mode and the $\phi$  Gauge Modes}
As shown in Appendix A, the collective fluctuations about the saddle
point is described by the quadratic form

\be
S_{\mbox{\rm eff}} = X_\alpha \left(
M_{\alpha\beta}^J + M_{\alpha\beta}^F \right)
X_\beta, \rule{.09in}{0in}\alpha = 1...12
\en
where $X_\alpha = \left(  \delta U_\mu^\ell , ... , \delta R, a_0^1,
a_0^2, a_0^3
\right)$ are the 12 degrees of freedom  and $M_{\alpha\beta}^F (\bbox{q},
i\omega_m)$ are made up of fermion bubbles computed in Appendix A.
Here we study
the spectral functions

\be
S_A (\bbox{q},\omega) = Im C_A (\bbox{q}, i\omega_m \rightarrow \omega + i\eta)
\en
where

\be
C_A (\bbox{q}, i\omega_m) = \langle A (\bbox{q}, i\omega_m) A
(\bbox{q},-i\omega_m)
\rangle
\en
and we focus on three cases, the $\theta$ mode $(A = \theta_u)$, the
transverse and longitudinal $\phi$ gauge modes
$(A=v_{i,i+\mu})$.  Using Eq.(44) these are readily expressed in terms of

\be
\langle X_\alpha X_\beta   \rangle = \left( M^J + M^F \right)^{-1}
\en
and computed numerically.   We use parameters in Eq. (21) 
${\tilde t}/{\tilde J} = 16/3$, i.e.,  ${\tilde t}/J = 2$
and $t^\prime /t = -0.5$, which gives realistic fermion bandstructure
and we show results for $x = 0.1$.  The mean field parameters are
$\chi_0 = 0.376$ and $\Delta_0 = 0.255$.  Note
that the maximum energy gap is $E_g = \Delta_{\bbox{k} = (\pi,\pi)} =
4 \Delta_0 \tilde{J}$ We measure energy in
units of
$\tilde{J}$ and the lattice constant is set to unity.

It is important to note that we define the correlators in terms of
the eigenvectors [(Eq.44)] which are the
eigenvectors fo 
r $\bbox{q}=(\pi,\pi)$ and $\omega_m=0$.   Away from
$(\pi,\pi)$, the overlap with the true eigenmode
is modified and other modes may mix in.  Furthermore, for the
$\phi$-gauge modes, the assignment in Eq.(44)
corresponds to ``polarization'' in the $\hat{x}$ and $\hat{y}$
directions, and are the appropriate longitudinal
and transverse modes only for $\bbox{q}$ along the $(\bbox{q}_x,\pi)$
direction.  Thus the numerical  results
shown here should be viewed as providing a guide for the behavior of
the modes near $(\pi,\pi)$.  In the next
section we will compute correlation functions which are
experimentally observable in the entire Brillouin
zone.

\begin{figure}[tbp]
\centerline{
\psfig{figure=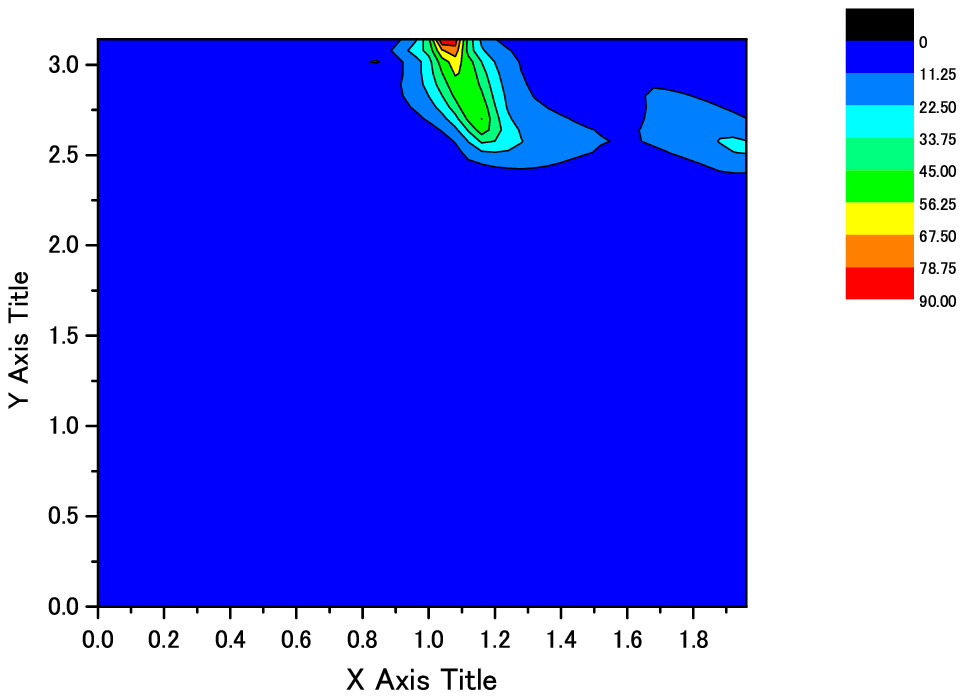,width=3.83in}
\psfig{figure=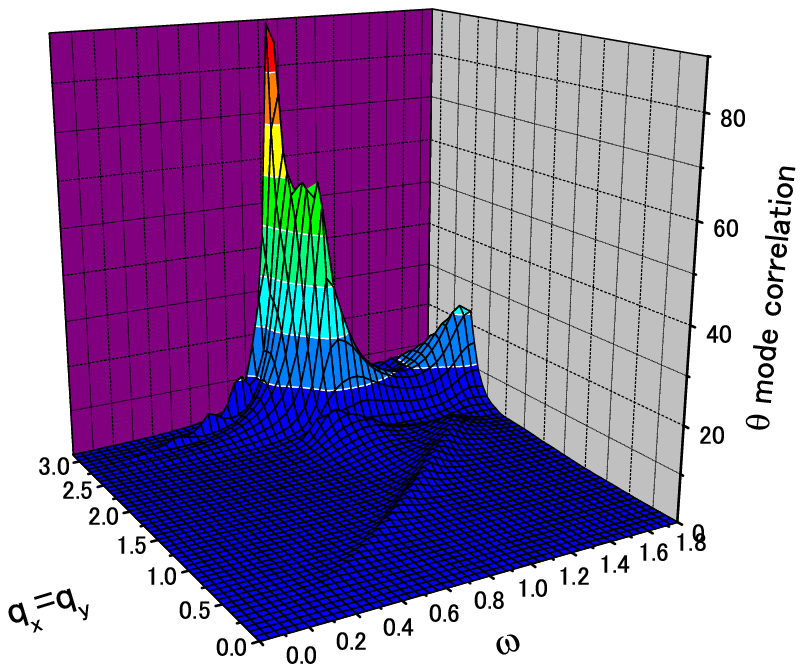,width=3.83in}
}
\caption{Spectral function for the $\theta$ mode for
$\bbox{q}=(q_x,q_y)$ from 0 to $(\pi,\pi)$.  A contour
plot  and a bird's eye view  are shown.  Frequency $\omega$ is in
units of $\tilde{J} (=
{3\over 8}J)$.
   }
\end{figure}

Figure 6 shows the results for the $\theta$ mode $S_\theta
(\bbox{q},\omega)$ with $\bbox{q}$ along the
diagonal $\bbox{q}_x = \bbox{q}_y$ in a contour plot and also in a
bird's eye view.  We see that the spectral
function shows a strong peak near $\omega \approx 1.1 \tilde{J}$
($\tilde{J}$ is defined after Eq.(8) and is
suggested to take the value $\tilde{J} = {3  \over 8} J$) which is
strongly localized near
$\bbox{q}=(\pi,\pi)$.  The spectral function disperses rapidly
upwards as $\bbox{q}$ deviates from
$(\pi,\pi)$.  The very large peak height can be anticipated from the
approximate form given by Eq.(45)
because the small value of the gap $ax$ gives a large spectral weight
upon inversion of the effective action.
The strong and narrow peak is responsible for the orbital current
fluctuations in the superconducting ground
state.\cite{16}

\begin{figure}[tbp]
\centerline{
\psfig{figure=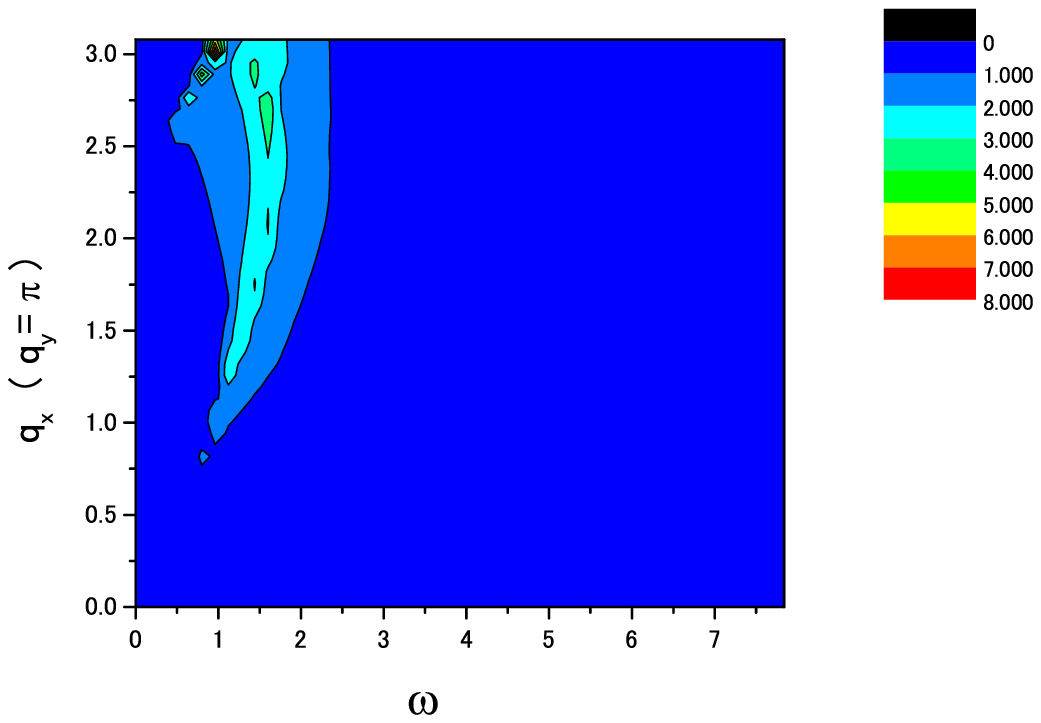,width=3.5in}
\psfig{figure=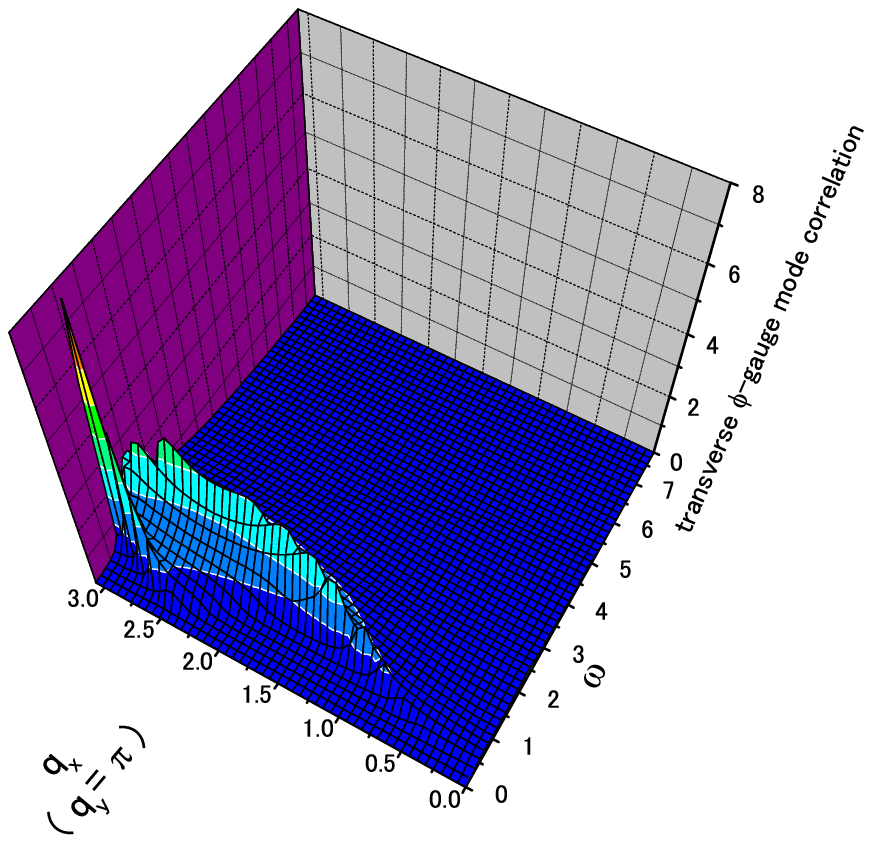,width=3.83in}
}
\caption{
Spectral function for the transverse 
$\phi$ gauge mode for $\bbox{q}=(q_x,\pi)$, 
i.e., from $(0,\pi)$ to $(\pi,\pi)$.
A contour plot  and a bird's eye view  are shown.  Frequency $\omega$ is in
units of $\tilde{J} (=
{3\over 8}J)$.
   }
\end{figure}

Figures 7 and 8 show the transverse and longitudinal $\phi$ gauge
modes along $\bbox{q}=(q_x,\pi)$.  The operators
are chosen to be

\be
\delta\phi_t = \cos \Phi_0 \delta\Delta_y - \sin \Phi_0 \delta\chi_y
\en

\be
\delta\phi_\ell = \cos \Phi_0 \delta\Delta_x + \sin \Phi_0 \delta\chi_x
\en

where

\be
\delta\chi_\mu = {1 \over 2}
( \sum_{\sigma} f^\dagger _{i \sigma} f_{i + \mu
\sigma} + c.c. ) - \chi_0
\en

\be
\delta\Delta_\mu = { 1 \over 2}
( f_{i\uparrow}f_{i +\mu\downarrow} - f_{i\downarrow} f_{i +\mu \uparrow}
 + c.c. ) - \Delta_0 \eta_\mu
\en
are the fluctuations of the real part of $\chi_{ij}$ and
$\Delta_{ij}$, respectively.  Since $\chi_0$, $\Delta_0$
are real, these correspond to amplitude fluctuations.  These
correlators are readily related to the correlations
involving $U_\mu^3$ and $U_\mu^1$, respectively.  Equations (56,57)
are simply the resolution of the fluctuations
illustrated in Fig.2 into its components along the vertical and
horizontal axes.
   Near
$(\pi,\pi)$ the two modes are degenerate and show a peak around
$\omega=1.5 \tilde{J}$.  The lineshape is
shown in Fig.9.  The mode is damped by particle-hole excitations and
a sharp feature appears as the frequency
drops below the particle-hole continuum.  (Recall that the 
particle-hole continuum is set by the
scale $\omega = 2E_g \approx 2\tilde{J}$ with our parameters.)  The
transverse mode frequency remains low  but loses spectral weight as $\bbox{q}$
goes away from ${\pi,\pi}$ towards $(0,\pi)$.  On the other hand, the
longitudinal mode disperses upwards and
gains in strength.

\begin{figure}[tbp]
\centerline{
\psfig{figure=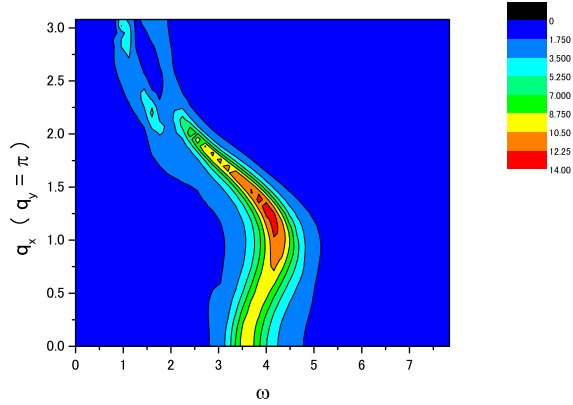,width=3.5in}
}
\caption{Contour plot of the spectral function for the longitudinal
$\phi$ gauge mode for
$\bbox{q}=(q_x,\pi)$, i.e., from $(0,\pi)$ to $(\pi,\pi)$.  Frequency
$\omega$ is in units of $\tilde{J} (=
{3\over 8}J)$.
   }
\end{figure}

\begin{figure}[tbp]
\centerline{
\psfig{figure=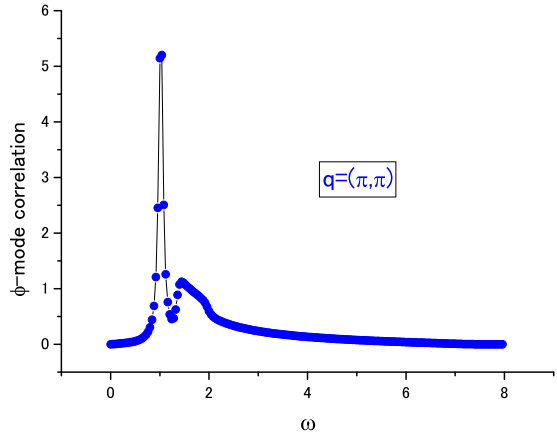,width=3.255in}
}
\caption{The lineshape for the  $\phi$ mode at
$\bbox{q}=(\pi,\pi)$.  Frequency $\omega$ is in units
of
$\tilde{J} (= {3\over 8}J)$.
   }
\end{figure}

Finally we discuss another important mode which is closely related to
the $\phi$ gauge mode.  It is the fluctuation
of the amplitude
$A_\mu = \sqrt{| \chi_{i,i+\mu} |^2 + | \Delta_{i,i+\mu} |^2 }$
and is parametrized by

\be
\delta A_\mu = \sin\Phi_0 \delta\Delta_\mu + \cos\Phi_0 \delta\chi_\mu \,\,\, .
\en
Unlike the modes discussed so far, the amplitude mode has finite
eigenvalue in the $x=0$ limit.  However, for $x
\approx 0.1$ the $\omega_m=0$ eigenvalue for the $\phi$ mode has
reached the value 0.493, quite close to the
amplitude mode eigenvalue of 0.669.  As seen from Eqs.(56,57,60) both
these modes involve the fluctuation
of  the amplitudes of $\chi$ and $\Delta$, and they will admix for
finite frequency.  This
complicates the interpretation of these collective modes.  In
contrast, the eigenvalue of the
$\theta$ mode is quite low at 0.155 and its interpretation as a
collective mode is more clear.  In
Fig.10 we show the spectral function for the ampltidue mode at
$\bbox{q} = (\pi,\pi)$.

\begin{figure}[tbp]
\centerline{
\psfig{figure=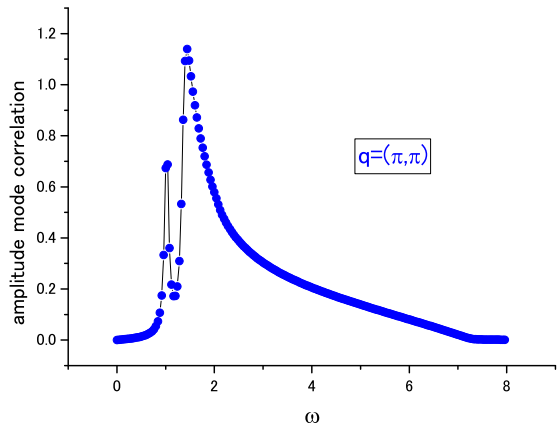,width=3.255in}
}
\caption{Spectral function of the amplitude mode at $\bbox{q} = (\pi,\pi)$.
   }
\end{figure}

\section{Experimental Observation of the Collective Modes}
Finally, we discuss possible experimental consequences of these
collective modes.  As discussed earlier, the $\theta$ mode
produces staggered orbital currents.  These in turn generate a
physical magnetic field which is staggered and which is, in
principle, observable by inelastic neutron scattering.  Here we
compute the scattering cross-section.  The neutron scattering
by orbital currents has been considered by Hsu {\it et al}.\cite{30} and we
follow their discussion.  We find that

\be
{d\sigma \over d \Omega d {\bbox{q}}} = {k^\prime \over k} \left(
{m_n\over 2\pi}  \right)
\left( {4\pi \mu_z \over Vq^2}    \right)^2 I  (\bbox{q},\omega)
\en

where

\be
I(\bbox{q},\omega) &=& {2\over 1-e^{-\beta\hbar\omega}} \left\{
\left( {q_y  \over q_x}  \right)^2 (1-\cos q_x) Im \chi_{x,x} (\bbox{q},\omega)
\right. \nonumber \\
&& + \left( {q_x\over q_y}  \right)^2 (1-\cos q_y) Im \chi_{y,y}
(\bbox{q},\omega) \nonumber \\
&& - {1\over 2} \left( 1-e^{iq_x} \right) \left( 1-e^{-iq_y} \right)
Im \chi_{x,y}  (\bbox{q},\omega) \nonumber \\
&& \left.\rule{-.05in}{.25in}
-{1\over 2} \left( 1-e^{-iq_x} \right) \left( 1-e^{iq_y} \right) Im
\chi_{y,x}  (\bbox{q},\omega)
\right\}
\en
where $\chi_{\mu\nu} = \langle J_\mu J_\nu  \rangle$, $\bbox{\mu} =
-1.91 {e\hbar \over m_nc} \bbox{S}$ is the neutron magnetic
moment, $m_n$ is the neutron mass, and $V$ is the volume.  We take the current
operator to be the mean field expression $\left( b=b_0 =
\sqrt{x}  \right)$

\be
J_\mu = ixt \sum_{i\sigma} \left(  f_{i\sigma}^\dagger
f_{i+\mu\sigma} - c.c.  \right) \,\,\, .
\en

In order to compute the correlation function, $\chi_{\mu\nu}$, we
note that any bilinear fermion operator can be written in
terms of $\psi_j^\dagger \Delta U_{ij} \psi_i$ for a suitable
$\Delta U$ which in turn can be expanded according to Eq.(26) as
$\Delta U_{i,i+\mu} = \sum_{a=0}^3 \eta_\mu^a \tau^a$.  The
correlation function is then computed by treating $\eta_\mu^a$ as
source terms and then differentiate the effective free energy with
respect to $\eta_\mu^a$.  The source terms simply modify
the effective action  $S_{\mbox{\rm{eff}}}$ [Eq.(52)] by

\be
S_{\mbox{\rm{eff}}}^\prime = X_\alpha M_{\alpha\beta}^J X_\beta +
X_\alpha^\prime M_{\alpha\beta}^FX_\beta^\prime
\en
where $X_\alpha^\prime$ is obtained from $X_\alpha$ by $\delta
U_\mu^a \rightarrow \delta U_\mu^a + \eta_\mu^a$.  Upon
completing the square and integrating out $X_\alpha$, we find that
$\langle  O_\mu^aO_\nu^b  \rangle$ where $O_\mu^a =
\psi_{i+\mu}^\dagger \tau^a \psi_i$ is obtained from the appropriate
matrix element of

\be
M^F \left(  1-(M^J + M^F)^{-1}M^F  \right) \,\,\, .
\en
In this way the neutron scattering cross-section is evaluated
numericaly.  We expect that the scattering is predominantly
coupled to the $\theta$ mode and indeed the result is very similar to
the $\theta$ mode spectral function shown in Fig.6.
In Fig.11 we show the lineshape
$I(\bbox{q},\omega)$ given by Eq.(62) in the zero temperature limit
at $\bbox{q}=(\pi,\pi)$.
In order to
estimate the experimental feasibilty, we compare the total
cross-section (integrated overall $\bbox{q}$ and $\omega$) to
the scattering from a lattice of $S = {1\over 2}$ moments.  Hsu {\it et al.}
estimated the total cross-section to be about 1\% of that of
spin scattering and our results are in rough agreement.  The
experimental detection of this signal may be difficult because
a resonance in the spin scattering exists around 30 meV at $\bbox{q}
= (\pi,\pi)$ for underdoped cuprates.  This resonance is
very narrow in $\bbox{q}$ and $\omega$, but its integrated weight
near $(\pi,\pi)$ is also about 1\% of the total spin
scattering.  The scattering due to the $\theta$ mode is of comparable
total strength but more spread out in $\omega$, making it harder
to detect.  However, unlike the spin
fluctuation which is isotropic, the orbital currents give rise to an
effective moment which is perpendicular
to the $a$-$b$ plane.  Since neutron is sensitive only to the
comonent of the moment normal to the $\bbox{q}$
vector, the orbital contribution can in principle be extracted by
varying the $\bbox{q}$ vector from normal to
parallel to the $a$-$b$ plane.

\begin{figure}[tbp]
\centerline{
\psfig{figure=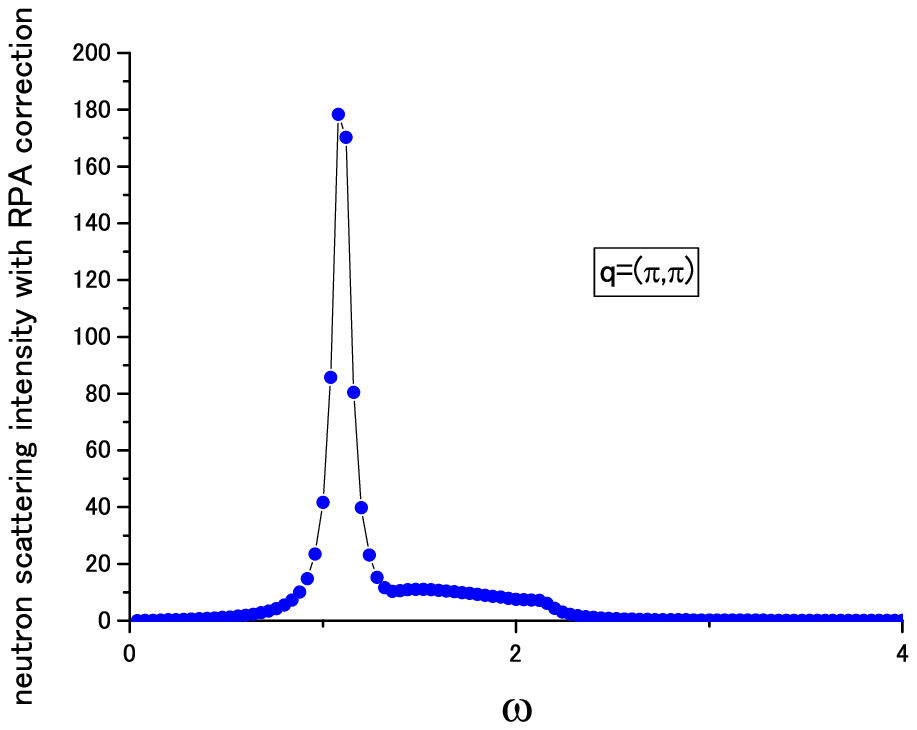,width=3.255in}
}
\caption{Neutron scattering intensity $I(\bbox{q},\omega)$ at zero
temperature vs. $\omega$ (in units of $\tilde{J}= {3\over
8}J$) at
$\bbox{q}=(\pi,\pi)$.
   }
\end{figure}

Another probe which couples to the orbital current is Raman
scattering.  However, since the excitation is epected to peak at
$(\pi,\pi)$, the large momentum  transfer requires X-ray Raman
scattering.  It is known that X-ray couples to magnetic moments as
well as orbital moments.\cite{31}
     The moment due to the orbital current is weak and corresponds in
strength to roughly $0.1 \mu_B$.  While Bragg
scattering from an ordered spin moment has been observed,\cite{32}
inelastic scattering from a short-range order of such
a small moment is beyond the current capability of X-ray scattering.
We note, however, that polarization dependence is
a powerful tood to distinguish between spin and orbital contributions
in X-ray scattering.\cite{31}

    Next we discuss the possible measurement of the $\phi$ gauge modes.
As discussed before, these modes involve the modulation
of the {\it amplitude} of the hopping parameter $\chi_{ij}$ and the
pairing parameter $\Delta_{ij}$ in a staggered fashion,
i.e., with momentum $(\pi,\pi)$.   The amplitude fluctuations of the
superconducting order parameter is not easy to detect and
has been experimentally observed only in the special case of
superconductivity in a charge density wave system.\cite{33,34}  On the other
hand, the fluctuations in $\chi_{ij}$ couples to the quasiparticle
hopping matrix element.  In the mean field theory we can
write an effective coupling as

\be
H_1 = -xt\sum_{<ij>\sigma}e^{i(e/c)A_{ij}}f_{i\sigma}^\dagger
f_{j\sigma} + c.c.
\en
where we include $A_{i,i+\mu} = a A_\mu$ as the lattice version
of the electromagnetic field $\bbox{A}$.  Expansion of
Eq.(66) to first order in $A_{ij}$ yields the standard
$\bbox{j}\cdot\bbox{A}$ coupling. When expanded to second order we
obtain ${1\over 2}xt \left( {e \over c} \right)
^2A_{ij}^2\left( f_{i\sigma}^\dagger f_{j\sigma}+ c.c. \right)$.  This
gives rise to Raman scattering which is coupled to
fluctuations in
$\chi_{ij}$. This kind of coupling was discussed by Shastry and
Shraiman\cite{35} as an explanation of the continuum background due
to incoherent electronic
excitations.  Here we expect that Raman scattering will couple to the
transverse and longitudinal
$\phi$ gauge mode.  Physically, a modulation of $|\chi_{ij}|$ is a
modulation of the {\it
bond} charge density which should couple to Raman scattering.
Standard Raman scattering provides essentially zero momentum
transfer.  In order to couple to the $\phi$ gauge mode at
$(\pi,\pi)$ and to follow its
dispersion, X-ray Raman scattering will be needed.  The leading
contribution to inelastic X-ray
scattering originates from the ${e^2\over 2mc^2} \bbox{A}^2$ term in
the single particle Hamiltonian
and the scattering cross-section is usually written as\cite{36}

\be
{d^2 \sigma\over d\Omega d\omega} = \left( {d\sigma \over
d\Omega}\right)_{Th} S(q,\omega)
\en
where the Thomson cross-section is

\be
\left(  {d\sigma \over d\Omega}  \right)_{Th} = r_0^2
\left(
\hat{\epsilon}_1 \cdot \hat{\epsilon}_2
\right)^2
{\omega_2 \over \omega_1} \,\,\, ,
\en
$r_0 = e^2/mc^2$ is the Thomson radius, $\hat{\epsilon}_1,\omega_1$,  and
$\hat{\epsilon}_2,\omega_2$ are the incident and scattering
polarization vector and frequency,
respectively, and $S(q,\omega)$ is the Fourier transform of the
electron density-density correlation
function.  On a lattice, the corresponding matrix element comes from
the second-order expansion of
Eq.(66) and can be written as

\be
&& {1\over 2} \sum_{<i,j>}xt \left({e\over c}\right)^2 A_{ij}^2
\left( f_{i\sigma}^\dagger f_{j\sigma} + c.c. \right)
\nonumber
\\ && = {x\over 4} \left({m \over m^*}\right) \left({e^2\over
mc^2}\right) {1\over 2}
\sum_{i,\mu}\bbox{A}^2_\mu (r_i)\left( f_{i\sigma}^\dagger
f_{i+\mu\sigma} + c.c. \right)
\en
where the sum is over nearest neighbors and $ta^2 = {1\over
4m^\ast}$.  Apart from ${m\over
m^\ast}$, which is of order ${1\over 2}$, we see that the coupling is
similar to the continuum
theory, except that the number operator
$f_{i\sigma}^\dagger f_{i\sigma}$ is replaced by
${1\over 4}\left( f_{i\sigma}^\dagger f_{i+\mu\sigma} + c.c.
\right)$.  Furthermore,
a factor $x$ arises due to the
strong correlation.  We see from Eq.(69) that X-ray Raman scattering
directly couples to the
fluctuating in $\chi_{ij}$ and therefore to the $\phi$ gauge mode.
It seems that high resolution
inelastic X-ray scattering is a promising technique to observe the
appearance of the $\phi$ gauge
mode at low temperatures. The cross-section for X-ray scattering is
proportional to
$< | \gamma(\bbox{q},\omega) |^2 >$
where
\be
   \gamma (\bbox{q},\omega) = \sum_\mu \hat{\epsilon}^\prime_\mu
\hat{\chi}_\mu (\bbox{q},\omega)
\hat{\epsilon}_\mu \,\,\, ,
\en
$\hat{\epsilon}$ and $\hat{\epsilon}^\prime$ are the incoming and
outgoing photon polarization and $\hat{\chi}_\mu
(\bbox{q},\omega)$ is the Fourier transform of the kinetic energy
operator defined in Eq.(58).
Thus the Raman scattering is given in terms of

\be
S_{\mu\nu}^\chi (\bbox{q},\omega) = < \hat{\chi}_\mu
(\bbox{q},\omega) \hat{\chi}_\nu (\bbox{q},\omega) >
\,\,\, .
\en
This is computed numerically using the same method described by
Eqs.(64) and (65).  The results are shown
in Fig. 12  along  $(q_x,\pi)$.
\begin{figure}[tbp]
\centerline{
\psfig{figure=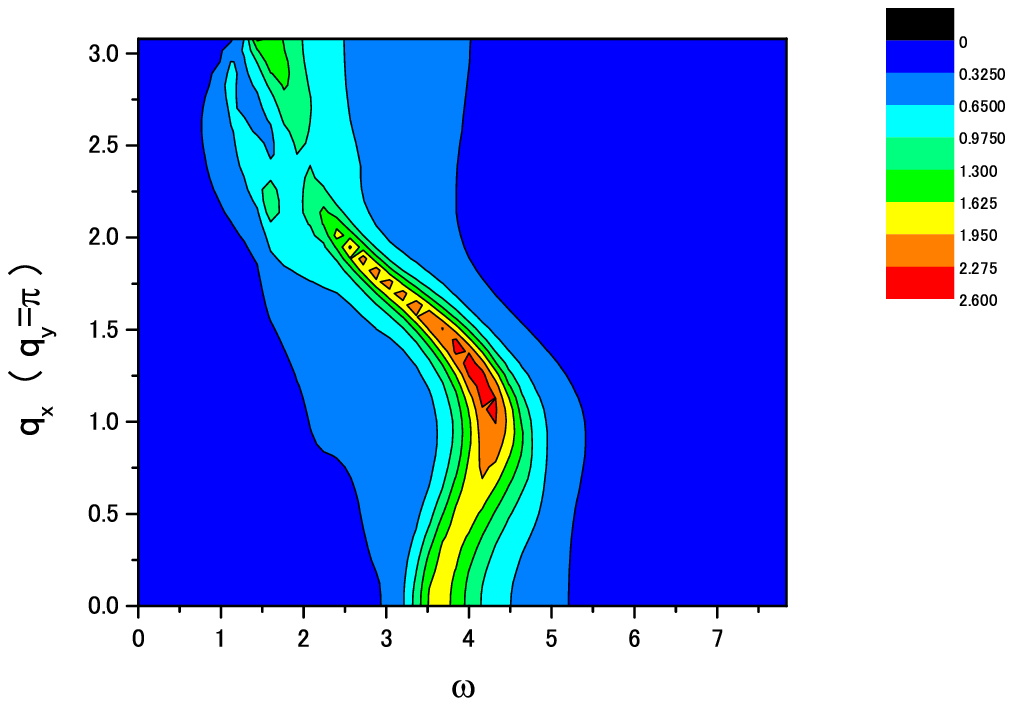,width=3.83in}
\psfig{figure=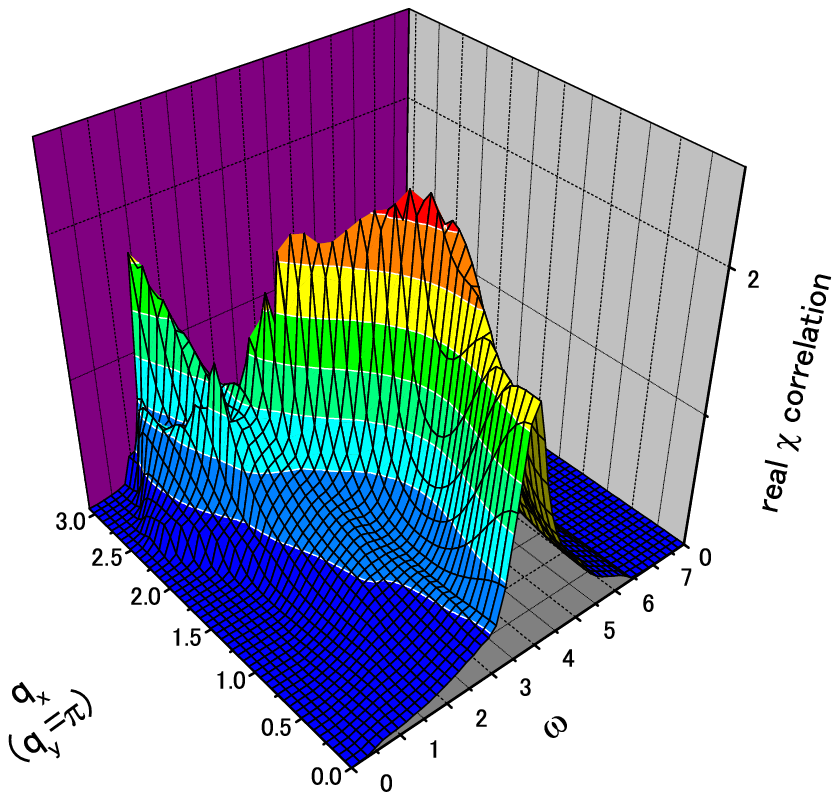,width=3.83in}
}
\caption{Spectral function
$S_{\mu\mu}^\chi(\bbox{q},\omega)(\mu=\hat{x})$ for the correlation
of the
kinetic energy operator [Eq.(71)] which is measured by inelastic
X-ray scattering.  A contour plot and a
bird's eye view are shown.
$\bbox{q}$ ranges from $(0,\pi)$ to $(\pi,\pi)$, i.e.,
$\bbox{q}=(q_x,\pi)$.  Frequency $\omega$ is in units of
$\tilde{J} (= {3\over 8} J)$.
   }
\end{figure}


The spectral function is peaked at
$(\pi,\pi)$ as expected.  Note that along the
$(\bbox{q}_x,\pi)$ direction (Fig.12) the
spectral function has similar structure as the longitudinal $\phi$
gauge mode (Fig.8).

The spectral function at $\bbox{q} = (\pi,\pi)$ is shown in Fig.13.
We note that its shape is quite similar to that of the amplitude mode
shown in Fig.10.  For
completeness, we also computed the correlation
$S_\Delta = < \delta\Delta_\mu \delta\Delta_\mu >$
where $\Delta_\mu$ was defined in Eq.(59).  This describes the
fluctuations of the real part of
$\Delta_{ij}$ and would correspond to the conventional
superconducting amplitude mode.  As shown
in Fig.14, its shape is similar to that of the $\phi$ gauge mode.

\begin{figure}[tbp]
\centerline{
\psfig{figure=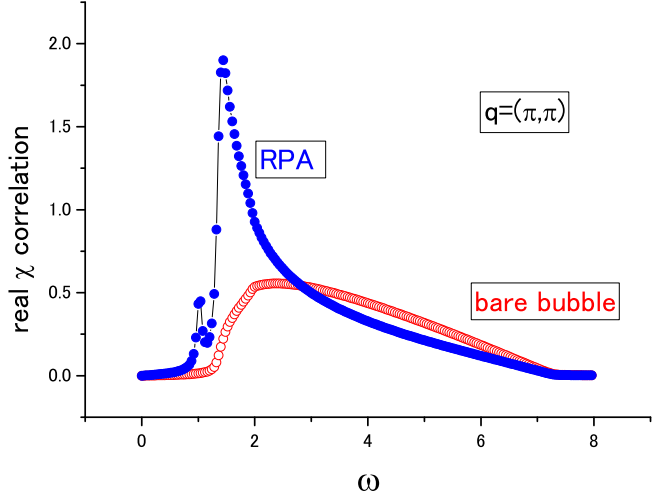,width=3.83in}
}
\caption{The lineshape of the kinetic operator correlator
$S^\chi_{\mu,\mu}(\bbox{q},\omega)$ $(\mu = \hat{x})$ as a function 
of $\omega$  (in
units of
$\tilde{J} = {3\over 8} J$) at $\bbox{q}=(\pi,\pi)$.  This is the 
curve labeled as RPA.  Also shown is the
curve labeled as bare bubble, which is computed by coupling to 
particle-hole excitations without allowing any
collective enhancement.
   }
\end{figure}

\begin{figure}[tbp]
\centerline{
\psfig{figure=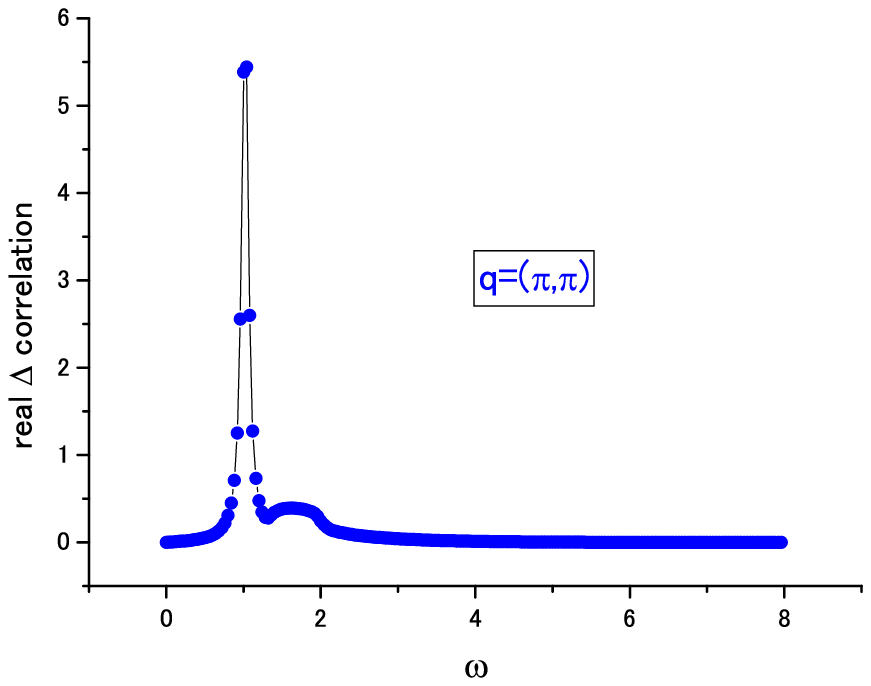,width=3in}
}
\caption{The spectral function $S_\Delta (\bbox{q},\omega)$ at
$\bbox{q} = (\pi,\pi)$. }
\end{figure}
Given that $\delta\chi$ and $\delta\Delta$ are linear superpositions
of $\delta A$ and
$\delta\phi$ according to Eqs.(56,60), it is surprising that their
spectral functions are not
simply weighted averages of the amplitude and $\phi$ gauge modes.
The reason is that there is
significant cross-correlation between $\delta\chi$ and
$\delta\Delta$, as well  as between
$\delta A$ and $\delta\phi$, i.e., the amplitude and $\phi$
fluctuations are not truly
eigenmodes.  This is a consequence of the significant mixing at
$x=0.1$ between these modes at
finite frequencies as discussed earlier.  It is perhaps better to
focus on $S^\chi_{\mu,\nu}$
which is experimentally measurable.  The peak observed in
$S^\chi_{\mu,\mu}$ corresponds to
collective oscillations of the hopping matrix element.  This is a new
degree of freedom not
present in conventional superconductors.
To emphasize this point we have also computed the ``bare bubble'' 
response, i.e., using $M^F_{\alpha\beta}$
without the collective enhancement shown in in Eq.(65).  As seen in 
Fig.13, the latter is much more spread out
in frequency.  Finally we note that for the relatively high doping 
considered here, the
$\phi$ gauge mode appears to be closely related to the more conventional
superconductor
amplitude fluctuation $\delta\Delta_\mu$ (albeit at $\bbox{q} =
(\pi,\pi)$) and is difficult to
detect.

We note that since $\chi_{ij}$ couples to the quasiparticle
hopping matrix element, it in turn is coupled to phonons.  In
Appendix B we consider the possibility that the collective modes
may appear as phonon sidebands which may be detected optically.

\section{Conclusion}
We have shown that in the gauge theory description of the $t$-$J$
model, new collective modes appear in the superconducting
ground state.  The SU(2) gauge theory allows us to classify these
collective modes and predict them analytically.  These
predictions are confirmed by numerical computation of the collective
mode spectra.  In particular, we see that the SU(2)
formulation allows us to smoothly connect to the $x=0$ limit, whereas
in the U(1) formulation, spurious modes would have
appeared.

In this paper we describe the bosons in the radial gauge.  It is then
clear that boson dynamics do  not play a role and
     that collective modes dynamics are entirely determined by the
fermions via the fermion bubbles which make up the matrix
$M^F$.  In Ref. (24) we proposed a $\sigma$-model formulation
where the low-lying excitations are parametrized by the boson
$
h = \sqrt{x}z = \sqrt{x}
\pmatrix{z_1 \cr
z_2}
$.
We proposed an effective action which included a term $x z^\dagger
D_0 z$ where $D_0 = \partial_\tau + eA_0 + a_0^3 \tau^3$
[Eq.(78) Ref. (24)].  Note that this term produces a
coupling between the $\theta$ and $\phi$ modes of the form
$\delta\theta\partial_\tau \phi$.  This Berry's  phase term is absent
in the present paper.  Thus Eq.(78) of ref. (24)
disagrees with the present work and we believe that it is incorrect.
Within the $\sigma$-model approach in ref. (24), one
should have absorbed the $z^\dagger\partial_\tau z$ term into the
$z^\dagger a_0^3 \tau^3 z$ term by a gauge transformation and
solve for the $a_0^{\prime\ell}$ terms by imposing the constraint
locally.  That is in fact what is done in this paper for
small $\delta\theta$ and $\delta\phi$.  A similar procedure can be
adopted for arbitrary $\theta$ and $\phi$ to generate a
$\sigma$-model.   Thus the $\sigma$-model should be viewed as a way
to parametrize the low-lying fluctuation of the $U_{ij}$
matrix [as in Eq.(40)] and the effective action depends only on the
fermion dynamics.

The appearance of the new collective modes answers the question of
whether the superconducting state described by the gauge
theory is any different from a conventional  BCS state.  The $\theta$
mode is coupled to staggered orbital currents.  The
importance of these currents was already revealed in the Gutzwiller
projected BCS wavefunction.\cite{16}  In principle, they should be
observable by neutron scattering or X-ray Raman scattering. A new
collective degree of
freedom is the moduation of the hopping amplitude which is observable
by high resolution
inelastic X-ray scattering.  We also show how they give rise
to side-bands in certain phonon modes.  However,  the weakness of the
coupling makes
its observation difficult.  The predictions of these collective modes are
unique features of the slave-boson/gauge-field approach to the $t$-$J$
model and experimental searches for the collective modes will serve
as important tests of this line of approach to the high T$_c$ problem.

\acknowledgements
We are thankful to Carsten Honerkamp, Xiao-Gang Wen and Jan Zaanen for
discussions.  P.A.L. acknowledges support by NSF grant number
DMR-0201069.

\appendix
\section{Derivation of the effective action for collective modes}
In this appendix, we sketch the calculation of the effective action
for the collective modes up to quadratic order.
For simplicity, we consider here the case of $t'=0$.
The inclusion of $t'$-term is rather trivial.
We start with the Lagrangian Eq.(21) in the text,
because we consider the superconducting ground state with the bose
condensation. The procedure is standard as in the usual 1/N expansion. Namely
we first divide the integral variables into the mean field (saddle point)
values and the fluctuations as given in Eq.(24).
Integrating over the fermionic integral variables, we can expand the
effective action with respect to the fluctuating part
$ \delta U_{ij}$, $\delta a_0^3$, $\delta a_0^1, \delta a_0^2$,
and $\delta R$.

Then the stationary condition that the linear
order terms in $\delta U_{ij} $ etc. vanish gives the self-consistent
mean field equations as
\begin{equation}
\chi = \int_{- \pi}^\pi { {d k_x} \over { 2 \pi} }
\int_{- \pi}^\pi { {d k_y} \over { 2 \pi} }
{ { - \xi_k \gamma_k} \over {2 E_k} }
= \int  {d^2 \bbox{k} \over (2\pi)^2}  { { - \xi_k \gamma_k} \over {2 E_k} }
\end{equation}
\begin{equation}
\Delta = \int {d^2 \bbox{k} \over (2\pi)^2}
{ { {\tilde J} \Delta \beta_k^2} \over { E_k} }
\end{equation}
\begin{equation}
\lambda_0 = \int {d^2 \bbox{k} \over (2\pi)^2}
{ { 2 {\tilde t} \gamma_k \xi_k } \over { E_k} }
\end{equation}
where $\gamma_k = \cos k_x + \cos k_y$,
$\beta_k = \cos k_x - \cos k_y$, and
$E_k = \sqrt{ \xi_k^2 + \Delta_k^2 }$ with $\xi_k$ and $\Delta_k$ being
given below eq.(23).

The Gaussian fluctuations are represented by the quadratic terms
$S_2$.
\begin{eqnarray}
S_2 &=& \int_0^\beta d \tau
\biggl[    {\tilde J} \sum_{i, \mu=x,y}
\bigl( |\delta \chi_{i i + \mu} |^2+|\delta \Delta_{i i + \mu} |^2
\bigr)
- r_0^2 \sum_i \bigl(
    2 i \delta a^3_0(i) \delta R_i + \lambda_0 (\delta R_i)^2 \bigr)
\nonumber \\
&-& \sum_{<i,j> \sigma} r_0^2 {\tilde t}_{ij} \delta R_i \delta R_j
< f^\dagger_{i \sigma} f_{j \sigma} + f^\dagger_{j\sigma} f_{i \sigma} >
\biggr]
\nonumber \\
&+& { 1 \over 2} Tr [G V G V]
\end{eqnarray}
where $\beta$ is the inverse temperature and
the last term represents the second order contributions from
the fermionic determinant.
Here the Green's function is $2 \times 2$ matrix for each
$k = ({\bbox{k}},\omega_n)$ (${\bbox{k}}$: wave vector, $\omega_n$:
fermionic Matsubara frequency).
Then the explicit form of the last term in eq.(A4) is
($q = ({\bbox{q}},\omega_m)$)
\begin{equation}
{ 1 \over { 2 \beta V} }\sum_q \sum_k
tr[ G(k-q/2) V(k-q/2,k+q/2) G(k+q/2) V(k+q/2,k-q/2) ],
\end{equation}
where $tr$ means the trace over the $2 \times 2$ matrix.
The Green's function is given by,
\be
[G(k)]^{-1} =
\pmatrix{i\omega_n - \xi_{\bbox{k}} &,& -\Delta_{\bbox{k}} \cr
-\Delta_k &, &  i\omega_n + \xi_{\bbox{k}} },
\en
which is represented in a compact form
in terms of the Pauli matrices $\tau^1, \tau^2, \tau^3$ as
\begin{equation}
G(k) = -[ \omega_n^2 + E_k^2 ]^{-1}
[ i \omega_n + \Delta_k \tau^1 - \xi_k \tau^3 ].
\end{equation}
The matrix $V$ is explicitely given by
\begin{eqnarray}
& & V(k+q/2,k-q/2)= { {2 {\tilde J}} \over { \sqrt{ \beta V} } }
\sum_{\mu = x,y}
\nonumber \\
& \times &
\pmatrix{ -i \delta a_0^3(q)
- \delta \chi''_\mu (q) \sin k_\mu
- \delta \chi'_\mu (q) \cos k_\mu &,&
    -i \delta a_0^1(q) - \delta a_0^2(q) +  \delta \Delta'_\mu(q) \cos k_\mu
+ i\delta \Delta''_\mu(q) \cos k_\mu \cr
-i \delta a_0^1(q) + \delta a_0^2(q) +
\delta \Delta'_\mu(q) \cos k_\mu
- i\delta \Delta''_\mu(q) \cos k_\mu &, &
    + i \delta a_0^3(q)
- \delta \chi''_\mu (q) \sin k_\mu
+ \delta \chi'_\mu (q) \cos k_\mu }.
\end{eqnarray}
Here it is noted that  $\delta U_{ij}$ is represented as
\begin{equation}
\delta U_{i i+\mu} = - \delta \chi'_\mu(i) \tau^3 + i \delta \chi''_\mu(i)
+ \delta \Delta'_\mu(i) \tau^1 - \delta \Delta''_\mu(i) \tau^2
\end{equation}
where $\chi_\mu(i) = \chi'_\mu(i) + i \chi''_\mu(i)$, and
$\Delta_\mu(i) = \Delta'_\mu(i) + i \Delta''_\mu(i)$.
The Fourier transformations of these variables are defined as
\begin{equation}
\chi'_\mu(i,\tau) = {1 \over {\sqrt{\beta V} } }
\sum_q \chi'_\mu (q) e^{ - i \omega_m \tau + i{\bbox{q}} \cdot
{\bbox{R}}_{i + \mu/2} },
\end{equation}
and similar expressions for other variables.
We define the 12 variables $X_\alpha(q)$ ($\alpha=1,...,12$) as
\begin{equation}
X (q) =
\pmatrix{
\delta \chi''_x(q) \cr
\delta \chi''_y(q) \cr
\delta \Delta'_x(q) \cr
\delta \Delta'_y(q) \cr
\delta \Delta''_x(q) \cr
\delta \Delta''_y(q) \cr
\delta \chi'_x(q) \cr
\delta \chi'_y(q) \cr
\delta R(q) \cr
\delta a^1_0(q)    \cr
\delta a^2_0(q)    \cr
\delta a^3_0(q)}
\end{equation}
In the calculation of the fermionic polarization function the following
integrals are needed.
\begin{equation}
F_{ab}({\bbox{q}},\omega_m; {\bbox{k}}) =
{ 1 \over { \beta} } \sum_{\omega_n}
tr[ G(k+q/2) \tau^a G(k-q/2) \tau^b],
\end{equation}
where $\tau^0 = 1$ (unit matrix) and $\tau^{1,2,3}$ are
Pauli matrices.
At zero temperature, the summation over $\omega_n$ is reduced to the
integral, i.e., $\beta^{-1} \sum_{\omega_n} \to \int d \omega /2 \pi$,
which can be done to result in the following $4 \times 4$ matrix.
\begin{eqnarray}
&  & F({\bbox{q}},\omega_m) =
\nonumber \\
&  & \left( \matrix{
(\omega_m^2/4+\chi_+ \chi_- + \Delta_+ \Delta_- ) I_1 - I_3,&
i(\Delta_+ + \Delta_-) I_2 + i \omega_m (\Delta_- - \Delta_+) I_1/2,\cr
i(\Delta_+ + \Delta_-) I_2 + i \omega_m (\Delta_- - \Delta_+) I_1/2,&
(\omega_m^2/4- \chi_+ \chi_- + \Delta_+ \Delta_- ) I_1 - I_3,\cr
i ( \chi_+ \Delta_- - \chi_- \Delta_+ ) I_1,&
(-\chi_+ + \chi_-) I_2 + \omega_m (\chi_+ + \chi_-) I_1/2,\cr
-i(\chi_+ + \chi_- )I_2 + i \omega_m (\chi_+ - \chi_- ) I_1/2, &
- ( \chi_+ \Delta_- + \chi_- \Delta_+ ) I_1, \cr}   \right. \nonumber
\end{eqnarray}
\begin{eqnarray}
& & \rule{1.90in}{0in}
\left. \matrix{
i ( -\chi_+ \Delta_- + \chi_- \Delta_+ ) I_1,&
-i(\chi_+ + \chi_- )I_2 + i \omega_m (\chi_+ - \chi_- ) I_1/2 \cr
(\chi_+ - \chi_-) I_2 - \omega_m (\chi_+ + \chi_-) I_1/2,&
- ( \chi_+ \Delta_- + \chi_- \Delta_+ ) I_1 \cr
(\omega_m^2/4- \chi_+ \chi_- - \Delta_+ \Delta_- ) I_1 - I_3,&
(-\Delta_+ + \Delta_-) I_2 + \omega_m (\Delta_+ + \Delta_-) I_1/2 \cr
(\Delta_+ - \Delta_-) I_2 - \omega_m (\Delta_+ + \Delta_-) I_1/2, &
(\omega_m^2/4 + \chi_+ \chi_- - \Delta_+ \Delta_- ) I_1 - I_3 } \right)
\end{eqnarray}
where
\begin{eqnarray}
I_1(q) &=& { 1 \over { \omega_m^4+ 2 \omega_m^2 ( E_+^2 + E_-^2) +
( E_+^2 - E_-^2)^2 } }
\biggl[
{ {\omega_m^2 - E_+^2 + E_-^2} \over { 2 E_+ } } +
{ {\omega_m^2 + E_+^2 - E_-^2} \over { 2 E_- } }
\biggr]
\nonumber \\
I_2(q) &=& { 1 \over { \omega_m^4+ 2 \omega_m^2 ( E_+^2 + E_-^2) +
( E_+^2 - E_-^2)^2 } }
\biggl[ { { \omega_m (E_+ - E_-) [ \omega_m^2 + (E_+ - E_- )^2] }
\over  { 4 E_+ E_-  } }
\nonumber \\
I_3(q) &=& { 1 \over { \omega_m^4+ 2 \omega_m^2 ( E_+^2 + E_-^2) +
( E_+^2 - E_-^2)^2 } }
\biggl[
{ { (E_+ + E_- )\omega_m^4 } \over { 8 E_+ E_-  } } +
{ { (E_+ + E_- )^3 \omega_m^2}  \over { 8 E_+ E_-  } } +
{ { (E_+ - E_- )^2 ( E_+ + E_-)}  \over { 2 } }
\biggr]
\end{eqnarray}
The static limit of these funcitons are easily estimated as
\begin{eqnarray}
I_1({\bbox{q}}, \omega_m=0)
&=& { 1 \over { 2 E_+ E_- ( E_+ + E_-)} }
\nonumber \\
I_2({\bbox{q}}, \omega_m=0) &=& 0
\nonumber \\
I_3({\bbox{q}}, \omega_m=0) &=& { 1 \over { 2 ( E_+ + E_- ) } }.
\end{eqnarray}
Here we have introduced the abbrebiations such as
$E_{\pm} = E_{ {\bbox{k}} \pm {\bbox{q}}/2 }$  etc.
Then the quadratic action $S_2$ with respect to the variables
$X_\alpha$ is given by
\begin{eqnarray}
S_2 &=& \sum_q  \biggl[ \sum_{ \alpha , \beta =1,12} \Pi_{\alpha \beta}(q)
X_\alpha(q) X_{\beta}(-q)
\nonumber  \\
&+& \sum_{\alpha = 1,8}  {\tilde J}
    X_\alpha(q) X_{\alpha}(-q)
- i x ( X_{12}(q) X_{9} (-q) + X_{12}(q) X_{9} (-q) )
- { { x \lambda_0} \over 2 } ( \cos q_x + \cos q_y ) X_9(q) X_9(-q).
\biggr]
\end{eqnarray}
Here
\be
\Pi_{\alpha \beta}
= { 1 \over V} \sum_k \zeta_\alpha (k) \zeta_\beta (k)
F_{a(\alpha),a(\beta)} ({\bbox{q}},\omega_m; {\bbox{k}})
\en
where
\be
\zeta(k) =
\pmatrix{
- 2 \sin k_x \cr
- 2 \sin k_y \cr
     2 \cos k_x \cr
     2 \cos k_y \cr
- 2 \cos k_x \cr
- 2 \cos k_y \cr
- 2 \cos k_x \cr
- 2 \cos k_y \cr
    - 4 x {\tilde t}(\cos(k_x) \cos(q_x/2)
       + \cos(k_y) \cos(q_y/2) )   \cr
- 1    \cr
- 1    \cr
- 1    }
\en
and $a(\alpha)$ is the index of the Pauli matrix
corresponding to each component and is
given as $a(1) = a(2) = 0, a(3) = a(4) = 1, a(5) = a(6) = 2,
a(7) = a(8) = a(9) = 3, a(10) = 1, a(11) = 2, a(12) = 3$. The fermion
bubble is analytically
continued in the standard way as $F (i\omega_m \rightarrow \omega +
i\eta)$. Equation (A16) is
numerically evaluated by discretizing the 1st Brillouin zone by $ 200
\times 200$  and treating
$\eta$
    as small but finite. The convergence with respect to the number of
the lattice points has been
checked.

Equation (A15) is the quadatic forms for the 12 variables $X_{\alpha}$,
and by integrating over the last 3 variables $X_{10},X_{11},X_{12}
( = \delta a_0^1, \delta a_0^2, \delta a_0^3)$,
we obtain the effective action for
9 variables, which has positive definite engenvalues for each
$q = ( {\bbox{q}}, \omega_m)$ when the mean field solution is stable.

\section{Coupling of the $\phi$ Gauge Mode to Phonons}
Let us consider the mode where the planar oxygen
moves in and out of the plane.  We focus on the oxygen mode
because the mass is light and the frequency relatively low, and both
features tend to enhance the coupling.  In LSCO and YBCO,
the oxygen is buckled out of the plane.  (In YBCO the displacement
$u_0 = 0.256$ Angstroms.)  Then the out-of-plane phonon mode
has a linear coupling to the Cu-O bond length and therefore to the
effective hopping $t$ and exchange $J$.  This problem
was considered by Normand {\it et al.}\cite{37} who concluded that in YBCO

\be
\delta t_{ij} &=& \lambda_t t \delta u_{ij}/a \\
\delta J_{ij} &=& \lambda_J J \delta u_{ij}/a
\en
where they estimate $\lambda_t \approx {1\over 2} \lambda_J \approx
2.6$ for a displacement $\delta u_{ij}$ of the oxygen on
the $ij$ bond normal to the plane.  The surprising large coupling is
partly due to the fact that the displacement $\delta
u_{ij}$ is normalized to the lattice constant $a \approx 4$ Angstroms  which is
quite large.  Let us first focus on the $\delta t_{ij}$ term.  The
modulation of the Hamiltonian is

\be
\delta H_t &=& \sum_{<ij>\sigma} \delta t_{ij} c_{i\sigma}^\dagger
c_{j\sigma} \nonumber \\
&\approx& \sum_{<ij>,\sigma} \lambda_t t {\delta u_{ij} \over a}
b_ib_j^\dagger f_{i\sigma}f_{j\sigma}^\dagger \nonumber \\
&\approx& \lambda_t (\chi_0 + \delta\chi_{ij}) xt{\delta u_{ij} \over a}
\en
The last line is in the mean field approximation,  where we retained
the amplitude fluctuation $\delta\chi_{ij}$ of $\chi_{ij}$.

Next, recall that the $\phi$ gauge mode couples to $\chi_{ij}$ via
Eqs.(37) and (38), so that

\be
\delta\chi_{ij} = -\Delta_0(-1)^{i_x+i_y} v_{ij} \,\,\, .
\en
Combining Eqs.(B3) and (B4), we find an effective coupling
between the pnonon displacement and the $\phi$ guage mode
co-ordinate

\be
H_{eff} &=& - \lambda_{eff} (-1)^{i_x+i_y} {\delta u_{ij} \over a} v_{ij}
\en
where

\be
\lambda_{eff} = \lambda_t x t\Delta_0 \,\,\, .
\en
Note that a fluctuation of $v_{ij}$ couples to a $(\pi,\pi)$
phonon mode as expected.  It turns out that the modulation
of $J$ given by Eq.(B2) does not couple to the $\phi$ gauge mode.
The reason is that in mean field, the modulation of
the
$J$ term is $\delta J_{ij} (\chi_{ij}^\ast \chi_{ij} + \Delta_{ij}^\ast
\Delta_{ij})$.  This couples to the amplitude mode.  On the other
hand, in the $\phi$ gauge mode,
the total modulus $|\chi_{ij}|^2 + |\Delta_{ij}|^2$ is held fixed, as
shown in Fig. 2.  Thus there is no coupling between the $\phi$ gauge
mode and the
phonon via the
$\delta J_{ij}$ term.

We next approximate the phonon as an Einstein model. The energy is
given by ${1\over 2}(-\omega^2 + \omega_0^2) a^2M \left(
{\delta u \over a} \right)^2$ where $\omega_0$ is the frequency and
$M$ is the mass.  We approximate the gauge mode as a well
defined mode at frequency $\omega_\phi$ with energy ${1\over 2}
{\alpha_\phi \over \omega_\phi} (-\omega^2 +
\omega_\phi^2) v_{ij}^2$ and adjust $\alpha_\phi$ to match the
spectral weight.  The coupled phonon-$\phi$ gauge modes are obtained by
diagonalizing the $2\times 2$ matrix

\be
D^{-1}=
\pmatrix{
{1\over 2}{\alpha_\phi \over \omega_\phi} (-\omega^2 + \omega_\phi^2)
& -\lambda_{eff} \cr
     -\lambda_{eff} & {1\over 2} (-\omega^2 + \omega_0^2)a^2M } \,\,\, .
\en
The modes are

\be
\omega_\pm^2 = {1\over 2} \left[
(\omega_0^2 + \omega_\phi^2) \pm \sqrt{(\omega_0^2 - \omega_\phi^2)^2
+ 4 g^2 \omega_0^4} \,\,
\right]
\en
where the dimensionless coupling constant is

\be
g^2 = {4\lambda_{eff}^2 \omega_\phi \over a^2M \omega_0^4 \alpha_\phi}
\en
The phonon Green's function $D = \langle \delta u_{ij}  \delta u_{ij}
\rangle$ is

\be
D = {(-\omega^2 + \omega_\phi^2) \over (-\omega^2 +
\omega_+^2)(-\omega^2 + \omega_-^2)a^2M} \,\,\, .
\en
Assuming $\omega_0 > \omega_\phi$, the phonon
mode is shifted slightly upwards in frequency to
$\omega_+$ but a side band appears at $\omega_-$ which is near
$\omega_\phi$.  The ratio of the spectral weight of the
$\omega_-$ side band to the $\omega_+$ mode is

\be
R = {\omega_\phi^2 - \omega_-^2 \over \omega_+^2 - \omega_\phi^2}
{\omega_+ \over \omega_-} \,\,\, .
\en
For weak coupling $g^2 << 1$, we expand the numerator using Eq.(B8)
and replace $\omega_+$ by $\omega_0$ and $\omega_-$ by $\omega_\phi$  to obtain

\be
R = {g^2\omega_0^4 \over \left( \omega_0^2 - \omega_\phi^2  \right)^2 }
\left( {\omega_0 \over \omega_\phi} \right) \,\,\, .
\en
This result for the relative spectral weight is also valid if
$\omega_0 < \omega_\phi$.

The spectral weight ratio is mainly determined by the coupling
constant $g^2$.  To make a rough estimate, we take $M =
\mbox{\rm oxygen mass} \,\, \omega_0 \approx J/4$ so that $a^2M
\omega_0^2 \approx 25 \,\mbox{\rm eV} \approx 200 J$.
Taking $\lambda_{eff} \approx J/4$, we find

\be
g^2 \approx {1\over 200} {\omega_\phi \over \omega_0\alpha_\phi} \,\,\, .
\en
We estimate $\alpha_\phi \approx 1$ because we can see from Fig. 9
that the spectral weight of the $\phi$ gauge mode at
$(\pi,\pi)$, is of order unity.  Unfortunately,
$g^2$ turns out to be very small.

The small spectral weight ($\sim 5 \times 10^{-3}$ of the main phonon
peak) means that it is probably impossible to observe
the side-band by neutron scattering.  Optical measurements offer
higher precision.  For YBCO the phonon is at
$\bbox{q}=(\pi,\pi)$ and do not couple to light.  LSCO offers a
special opportunity in that at low temperatures, the
lattice is in the low temperature orthorhombic phase (LTO) where the
buckeling of the oxygen is staggered, i.e., $u_{ij} =
u_0(-1)^{i_x+i_y}$. Then a $\bbox{q}=0$ distortion $u_{ij} = u_{ij}^0
+ \delta u_{ij}$ couples to the staggered
modulation in
$\chi_{ij}$.  Hence in LSCO the $\phi$ gauge mode is coupled to the
$\bbox{q}=0$ phonon mode which can be studied
optically either by absorption or by Raman, depending on its
activity.  The signature of the side band is that it should
appear only in the superconducting phase, because in the normal state
(pseudogap state) the  $\bbox{I}$ vector rotates
out of the plane and is disordered, so that the $\phi$ mode is
expected to be smeared out.  Similarly, the weight of the
side band will be reduced by applying a magnetic field by the
fraction of the sample occupied by the vortex core.  This
is because the $\bbox{I}$ vector is rotated to the north pole near
the vortex core and the $\phi$ mode will lose its
identity.

The above discussion  was based on the assumption that the $\phi$
gauge mode is a well defined
mode and that the amplitude mode $\delta A$ is much higher in
frequency and plays no role. This
is correct for small doping but we have seen that for $x=0.1$, the
numerical results show a
strong admixture of the $\phi$ gauge mode and the amplitude mode.
Nevertheless, it is still
correct that the phonon couples to $S^\chi_{\mu\mu}$, which produces
a sideband with a lineshape
given in Fig.13.  However we should  include the fact that the phonon
also modulates the exchange
constant according to Eq.(B2) and couples to the amplitude
fluctuation as well.  Since the
amplitude fluctuation has a very similar lineshape (see Fig.10) to
that of $S^\chi_{\mu\mu}$, the
final result may still be interpreted as a modulation  of the hopping
amplitude, a new collective
degree of freedom not present in conventional superconductors.
Finally, we remark that we only
have results for $T=0$, and in the $x=0.1$ case the issue of how much
of the spectral weight in
$S^\chi_{\mu\mu}$ survives above T$_c$ remains open.

\end{document}